\RequirePackage{fix-cm}
\documentclass[twocolumn]{svjour3}          
\smartqed  
\usepackage[utf8]{inputenc}
\usepackage{color}
\usepackage{graphicx}
\usepackage{bm}
\usepackage{multirow}
\usepackage{subfig}
\usepackage{xurl}
\usepackage{caption}
\usepackage [round]{natbib}
\usepackage{amsmath}
\usepackage[T1]{fontenc}
\usepackage{float}
\usepackage{epsfig}
\usepackage[]{lineno}
\usepackage{lscape}

\usepackage[utf8]{inputenc}
\usepackage{booktabs, makecell, longtable}

\begin{document}
\sloppy

\title{}

\title{ {Functional ANOVA approaches for detecting changes in air pollution during the COVID-19 pandemic}}

\titlerunning{Air pollution during the COVID-19 pandemic}        

%
%
%

\author{Christian Acal  \and Ana M. Aguilera \and Annalina Sarra \and Adelia Evangelista \and Tonio Di Battista \and Sergio Palermi}
\authorrunning{} 

\institute{C. Acal, A.M. Aguilera \at
              Department of Statistics and O.R. and IMAG, University of Granada (Spain) \\
              \email{chacal@ugr.es}, \email{aaguilera@ugr.es}         
               \and
         A. Sarra, A. Evangelista, T. Di Battista \at
             Department of Philosophical, Pedagogical and Economic-Quantitative Sciences, University G. d'Annunzio, V.le Pindaro, 42; 65127 Pescara (Italy) \\
             \email{asarra@unich.it}, \email{adelia.evangelista@unich.it}, \email{tonio.dibattista@unich.it}\\
             \and
           S. Palermi \at
              Agency of Environmental Protection of Abruzzo (ARTA), V.le G. Marconi, 51; 65127 Pescara  (Italy) \\
              \email{s.palermi@artaabruzzo.it}
}

\date{Received: date / Accepted: date}

\maketitle
\abstract{
 Faced with novel coronavirus outbreak, the most hard-hit countries adopted a lockdown strategy to contrast the spread of virus. Many studies have already documented that the COVID-19 control actions have resulted in improved air quality locally and around the world. Following these lines of research, we focus on air quality changes in the urban territory of Chieti-Pescara (Central Italy), identified as an area of greatest criticality in terms of air pollution. Concentrations of NO$_{2}$, PM$_{10}$, PM$_{2.5}$
 and benzene are used to evaluate air pollution changes in this Region. Data were measured by several monitoring stations over two specific periods: from 1st February to 10th March 2020 (before lockdown period) and from 11st March 2020 to 18th
 April 2020 (during lockdown period). The impact of lockdown on air quality is assessed through functional data analysis. Our work makes an important contribution to the analysis of variance for functional data (FANOVA). Specifically, a novel approach based on multivariate functional principal component analysis is introduced to tackle the multivariate FANOVA problem for independent measures, which is reduced to test multivariate homogeneity on the vectors of the most explicative principal components scores. Results of the present study suggest that the level of each pollutant changed during the confinement. Additionally, the differences in the mean functions of all pollutants according to the location of monitoring stations (background vs traffic), are ascribable to the PM10 and benzene concentrations for pre-lockdown and during-lockdown tenure, respectively. FANOVA has proven to be beneficial to monitoring the evolution of air quality in both periods of time. This can help environmental protection agencies in drawing a more holistic picture of air quality status in the area of interest.
}
\section{Introduction}\label{sec:intro}
After the discovery of the first case in Wuhan (China) in December 2019, the current outbreak of COVID-19, caused by severe acute respiratory syndrome coronavirus 2 (SARS-COV-2) has dramatically affected all the countries \citep{Wang2020}. On January 30 2020, the World Health Organization (WHO) declared worldwide public health emergency and in March 11 2020, due to widespread global infection, the WHO authorities categorised the new Coronavirus as pandemic \citep{WHO2020}.
To contain the virus and save lives, governments around the word have been taking a range of actions and measures, such as social and travel restrictions. More specifically, coronavirus pandemic has forced nations under partial or complete lockdowns, resulting in prohibition of unnecessary commercial activities in people's daily lives; prohibition of any types of gathering by residents; restrictions on private (vehicle) and public transportation.
Different studies have already documented the effects of COVID-19 lockdown measures on many aspects of human activities, such as transportation \citep{Mogaji2020},
renewable and sustainable energy \citep{Hosseini2020}, health risk assessment \citep{Gautam2020, Ambade2021, Zoran2020,Gupta2021}, tourism \citep{Sigala2020}, commodity markets \citep{Rajput2021}.
Certainly, COVID-19 has severe negative impact on the world activities as well as on local economy. In the major economics across the globe, lockdown will directly affect the Gross Domestic Product (GDP) of each country.
On the contrary, efforts to restrict transmission of the SARS-CoV-2 have had outstanding effects on the ecosystems which are being greatly recovered.
In many cities where lockdown measures have been implemented, the decline in economic activities, the non-functioning of industries, the drop in road transport, have contributed to mitigate air pollution.
\citet{Bherwani2020} monetarily quantified the overall benefit due to pollution reduction  over the potential economic loss sustained by local governments, as an outcome of lockdown.\\
Several researchers around the world reported that there is a considerable reduction of air pollution level across geographies.
For instance, the findings of many studies showed a substantial enhancement in the air quality in the lockdown period globally \citep{Lal2020,Dutheil2020,Gope2021,Venter2020,Gautam2020b}.\\
A visible improvement in air quality parameters was recorded in different areas of China and India (see, among others, \citet{Wang2020, Bao2020,Li2020,Gautam2020,Mahato2020,Sharma2020,Agarwal2020,Gautam2021}).
The study of \citet{Kanniah2020} highlighted that large emission reduction in transportation and anthropogenic activities, resulted in a significant reduction in air pollution levels in the urban regions of Malaysia as well as in a sizeable reduction aerosol optical thickness concentration during March-April 2020 in comparison with  the same period in 2019 and 2018.\\
\citet{Kerimray2020} analyzed the effect of the lockdown from March 19 to April 14, 2020, on the concentrations of air pollutants in Almaty, Kazakhstan, finding out
 a substantial difference between concentrations of air pollutant recorded before and during lockdown.\\
\citet{Otmani2020} gave evidence that the government decisions in response to COVID-19 had an important impact on the air pollution in Sal\'e city (Morocco). Findings from that study
showed that the difference between the concentrations recorded before and during the lockdown period were  75$\%$, 49$\%$ and 96$\%$ for PM$_{10}$, SO$_{2}$ and NO$_{2},$ respectively.\\
\citet{Berman2020} investigated the changes in levels of air pollutants across USA during COVID-19 pandemic. The authors reported a significant reduction on NO$_{2}$ (up to -25.5$\%$) and an overall decline in PM$_{2.5}$, compared with pre-lockdown phase.
\citet{Zabrano-Monserrate2020} studied the effects of quarantine policies on air pollutants concentrations in Quito, Ecuador. Using parametric methods, they detected a significant reduction of NO$_{2}$ and PM$_{2.5}$  since the introduction of lockdown measures. However, there was a noticeable growth in ozone levels.
There were reports that documented a decline in NO$_{2}$, NO$_{x}$ and an increase in surface O$_{3}$ at Sao Paulo in Brazil during the quarantine period
(see, for instance, \citet{Nakada2020,Dantas2020}).
The similar trends of reducing air pollution and increasing air quality due to introduction of lockdown were observed in large parts of Europe, such as France, Germany, Spain, and Italy \citep{Sicard2020,Collivignarelli2020, Zambrano-Monserrate2020,Tobias2020}.
An overview of selected studies on air pollution and COVID-19 is provided in the Appendix.\\
In this study, we focus on investigating the possible effects of the lockdown due to the COVID-19 pandemic on air quality in the Pescara-Chieti urban area, Abruzzo (Italy),
identified as an area of criticality in terms of air pollution.
Data of monitoring stations of the regional air quality network managed by the Regional Agency for the Environmental Protection (ARTA) of Abruzzo have been collected and examined. We compared data from  1$^{st}$ February to 10$^{th}$ March 2020, before the beginning of the main limitations on personal mobility, with data from  11$^{st}$ of March to  18$^{th}$ of April, during the adoption of lockdown restrictions. Measured concentrations of
NO$_{2}$, PM$_{10}$, PM$_{2.5}$ and benzene were used to evaluate air pollution changes.
Commonly, strategies used in monitoring air quality refer to descriptive statistics, box-plots, autocorrelation analysis and spatio-temporal models.
Unfortunately, in the monitoring of environmental pollutants, the temporal observations of the different pollutants  for the different stations have not always been referred to the same instants of time. As a result, the implementation of classical statistical procedures might be problematic.
Besides, for interpretative purposes, it is convenient to rely on statistical methods able to capture the speed and acceleration of pollutants variation over time.
For these reasons, in our research, to overcome the weakness of classical statistical procedures and to effectively  detect to what extent extreme changes in human behaviour after the quarantine policies adopted by the Italian Government have affected air quality, we followed an approach based on functional data analysis (FDA).
During the two  last decades, it has emerged an important literature in this methodological framework.
A comprehensive introduction to the foundations and applications of FDA can be found in \citet{Ramsay2002, Ramsay2005}, whereas nonparametric functional methods are summarized in a monograph by \citet{Ferraty2006}.
It is well known that  FDA extends the classical multivariate techniques to data whose observations are functions, usually curves, with the advantage of reducing a large number of discrete observations highly correlated for each curve to a functional form that conserves all  relevant information.

Recently, the use of FDA methods for environmental data has received attention.
For instance, \citet{Escabias200595}, proposed a functional principal component logistic regression to estimate the risk of drought in terms of time evolution in temperatures in Canada.
\citet{Gao2008} use functional methods to model the dynamics of diurnal ozone and nitrogen oxides cycles, their interrelationships, and the multilevel spatio-temporal variations of ozone and nitrogen oxides measurements from Southern California.
A functional model for forecasting the time evolution of a binary response from discrete time observations of a continuous time series, is introduced by \citet{Aguilera2008}
to predict the risk of drought in a future period of time from monthly observations of \emph{El Ni\~{n}o} phenomenon.
 \citet{Martinez2014} and \citet{MartinezTorres2020} implement a model based on functional analysis to detect outliers in air quality samples, with the overall aim to achieve a better solution for the air quality control. Likewise, \citet{Sancho2014} expand the concept of functional outliers to the set of control-chart techniques.
 Additional applications that prove the advantages of FDA in environmental research can be found in  \citet{Ocana2008,Valderra2010,Park2013,Shaadan2013,Escabias2013,Hormann2015,Agui2017,Gautam2020}.\\
Following these research streams, also in this study, rather than simply considering the data as vectors to apply classical multivariate analysis methods, which may lead to a loss of useful information, we explicitly exploit the functional form of environmental data.
It is worth noting that from a physical point of view the processes involved in the production of pollutants, such as NO$_{2}$ and particulate matter (PM$_{10}$, PM$_{2.5}$)
are of continuous time type and thus turn out to be appropriate for the analysis of functional data.
As a result, the use of models that account for the continuity of the whole trajectories along time seem to be more natural.
The FDA paradigm makes it possible to work with the entire time spectrum of pollutants time series.
In doing so, the FDA approach might bring additional information to be recovered from the data than in the vectorial approach, by looking at the smoothness
of underlying functions and its derivatives.
Furthermore, in contrast to most other methods commonly used to model trends in time series data, the functional techniques make non-parametric assumptions and there is no
concern about correlation due  to repeated measures.\\
Our goal in this paper is to ascertain whether the level of
each pollutant has changed during the lockdown period. In other terms, we want to test the equality of mean functions related to each pollutant in two different periods of time: before and during lockdown days.
The theoretical framework involves the use of FDA tools for repeated
measures, and in particular, the analysis of variance.
In the literature there are not many works related
to this matter for the field of FDA.
In this work, the statistics proposed by \cite{MartinezCamblor2011} and \cite{2020} to test the equality of two mean functions are extended by assuming a basis expansion of the sample curves. On the other hand, in order to check the differences between the temporal evolution of all pollutants in  terms of the location of measuring stations, a  novel approach for multivariate FANOVA for independent measures is introduced. This is based on multivariate  functional principal components analysis of the sample curves of all pollutants and the problem is reduced to testing multivariate homogeneity or MANOVA (gaussian data) on the vectors of the most explicative principal components scores.
 This new approach solves the problems of high dimension and multicollinearity that affect the basis expansion approach developed in  \citet{Gorecki2017} when the number of basis coefficients for each functional variable is large. The relevance of this research work lies in its contribution to extending the parametric and nonparametric functional homogeneity testing procedures proposed
by \citet{Aguilera2021} for  univariate functional data to the multivariate case.\\
In addition to this methodological development, this study provides empirical suggestions for local environmental authorities.
The comparison of air pollutants allows to study whether the level of each pollutant decreased during the quarantine days.
The suitable analysis of a different behaviour of pollutants at measuring sites provides ground to decide on whether to keep recording measurement at different
locations or redesign the local surveillance network, with the overall aim to reduce air pollution.\\
The paper is organized as follows.  Section \ref{sec:data} is dedicated to illustrate the studied area, the monitoring stations where air quality data have been collected and some explorative analysis
of pollutants used for this research. Section \ref{sec:thframe} introduces the theoretical framework.
In Section \ref{sec:multfan}, our method is applied for checking  differences between air quality data collected at
the monitoring stations placed in the urban area of Chieti-Pescara.
Finally, Section \ref{sec:conclu} concludes the paper.

\section{Air quality data and studied period}\label{sec:data}
This section gives details about the studied area with respect to its geographic characteristics and the monitoring stations used to collect air quality data. In addition, the dataset employed to derive insights into research problem has been analyzed with descriptive statistics and visualization tools. For high quality graphics, we used the \textit{`openair'} R package \citep{Carslaw2012}.

\subsection{Description of studied region}
\begin{figure}
	\includegraphics[width=8cm, height=6cm]{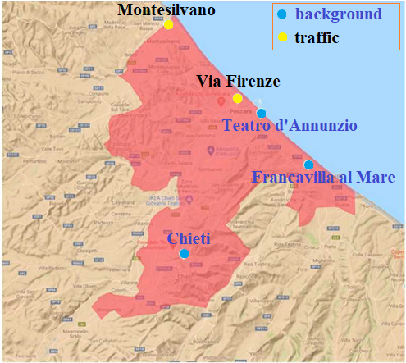}
	\caption{Abruzzo and Chieti-Pescara metropolitan area}
	\label{fig:metropol_area}
\end{figure}

In this work, we closely examine the air quality of the metropolitan area of Chieti-Pescara, situated in the Abruzzo region, along the Adriatic coast of central Italy. The Chieti-Pescara metropolitan area  (Fig.\ref{fig:metropol_area}) is a territory, identified according to a functional criterion, formed by six municipalities, Pescara, Montesilvano, Chieti, Francavilla al Mare, San Giovanni Teatino, Spoltore, covering a total area of 159.33 $km^2$, and accounting for around 281,101 inhabitants at 31/12/2019.

The configuration of Chieti-Pescara urban area is limited by the sea, in the North-East, and by hilly reliefs in the South-West. The central city is formed by the two provincial capitals: Chieti, not in a central position for the municipalities of the province, and the city of Pescara, which are extremely close to each other (approximately 12 km).
Pescara city, located on the centre of a metropolitan area (on the coast), is the administrative and commercial heart of Abruzzo and in a few decades, it has become the most populated city of the region, with 120,000 inhabitants. It developed on a flat territory, with a surface of 33.62 $km^2$, whose urban area develops around the terminal stretch of the homonymous river and a restricted coastal area.\\
The Chieti-Pescara conurbation is characterized by a system of infrastructures, which is one of the strongholds of the Abruzzo: significant and industrial sites are located around this pole. However, the progressive growth of the industrial activity, the increased road travel, the urban expansion, make the metropolitan area the locus of growing environmental concerns, for the rising levels of energy and resource consumption, greenhouse gas emissions and air quality pollution.
For these reasons, the conurbation of Chieti-Pescara has been identified as an area deserving mitigation measures to reduce air pollution by the "Plan for the Protection of Air Quality", drafted by the Abruzzo Region, in accordance with current Italian legislation (Legislative Decree 155/2010).
Also, in the Plan, there is the inventory of the main sources of polluting emissions (updated to 2012), which in this area largely sees the contribution of non-industrial combustion plants (mainly domestic heating plants) as regards particulate emissions (78.5$\%$ of the total for PM$_{10}$, 88.8$\%$ for PM$_{2.5}$ and 88.4$\%$ for benzene), while for nitrogen oxides road traffic prevails (49.7$\%$), with industrial combustion plants in second place with 23.6 $\%$.

\subsection{Data}
This study tracks four pollutants over two specific periods: from 1$^{st}$ February to 10$^{th}$ March 2020 (before lockdown period) and from 11$^{st}$ March 2020 to 18$^{th}$ April 2020 (strict lockdown period).
The analyses include measures of NO$_{2}$, PM$_{10}$, PM$_{2.5}$ and benzene obtained from the automatic reporting platform, run by Regional Agency for the Environmental Protection of Abruzzo Region (ARTA).
These variables are measured in micrograms per cubic meter ($\mu$g/m$^{3}$) and information are obtained from five monitoring sites. The spatial location of all five monitoring stations is shown by blue points in Figure \ref{fig:metropol_area}. The air monitoring stations of Pescara (Via Firenze) and Montesilvano are designed as \emph{Urban Traffic type} (UT) and are located roadside, where the pollution level is most influenced by traffic emissions from neighboring roads with medium-high traffic intensity; conversely, air quality data collected from the monitoring stations of Pescara (Teatro d'Annunzio), Chieti and Francavilla are deemed \emph{Urban Background} (UB), located where the pollution level is not influenced mostly by emissions from specific sources and are representative of the population average exposure.
Hourly measurements of pollutants have been collected from February to April 2020.

\subsection{Descriptive statistics and graphical analysis} \label{sec:descriptives}
After the implementation of strict lockdown measures starting from 11$^{st}$ March 2020, air pollution of the urban area of Chieti-Pescara has witnessed a substantial
improvement.
Table \ref{tab:mean} highlights the net and percentage variations of each pollutant, in each monitoring site, before and during the lockdown.
Early evidence of significant reduction of NO$_{2}$ (air pollutant mainly ascribable to fossil fuel combustion) as a consequence of lockdown, both in Asia and Europe, was showed by the data collected by NASA and ESA satellites \citep{Gautam2020a}. In our studied area it can be also noticed that NO$_{2}$ has shown the most significant declining trend. In particular, the concentrations of this pollutant were approximately $50\%$ lower compared to the previous average, that of pre-lockdown period. On the other hand, we recorded an increase of PM$_{10}$ and PM$_{2.5}$ concentrations during the lockdown weeks, whereas benzene levels dropped in the traffic measuring stations and increased in the background monitoring sites.
A trend analysis of 24-hour daily average data for the four pollutants was also considered for the above stated periods in all monitoring stations to better understand the impact in the levels of pollutants accumulation amid the lockdown period.\\

\begin{table}
	\small
	\caption{Net and $\%$ variation of pollutants concentration levels in the urban area of Chieti-Pescara}
	\vspace{0.3cm}
	\setlength{\tabcolsep}{6pt} 
	\begin{center}
		\begin{tabular}{ll|cc|ccc}
			
			&		&	\multicolumn{2}{c|}{\textbf{UT}}	&	\multicolumn{3} {c}{\textbf{UB}}	\\ \hline
			&	\textbf{Net variation} &	\textbf{fi}	&	\textbf{mo}		&	\textbf{th}		&	\textbf{ch}	&	\textbf{fr}	\\ \hline
			&	NO$_{2}$	&	-13.9	& -14.7	&	-21.2	&	-10.3	&	-7.6 \\
			&	PM$_{10}$ 	&	5.1	&	3.7 &	5.7	&	4.3	&	7.3	\\
			&   PM$_{2.5}$ 	&	2.9	&	2.2	&	3.1	&	4.4	&	4.1	\\
			&	Benzene	&	-0.31	&	-0.15	&	0.22	&	0.18	&	0.04	\\ \hline \hline
			& \textbf{$\%$ variation}  & & & & & \\ \hline
			&	NO$_{2}$	&	-57.9	& -58.7	&	-65.2	&	-54.8	&	-49.1 \\
			&	PM$_{10}$ 	&	20.5	&	16.8   &	22.0	&	19.4	&	40.8	\\
			&   PM$_{2.5}$ 	&	19.0	&	15.6	&	19.8	&	26.7	&	34.4	\\
			&	Benzene	&	-32.57	&	-27.56	&	40.06	&	19.63	&	4.27	\\	
			
		\end{tabular}

		\label{tab:mean}
	\end{center}
	\vspace{0.2cm}
	\small{
		\textbf{Acronyms of monitoring stations:}
		\vspace{0.1cm}\\
		fi=Via Firenze;
		mo=Montesilvano;
		th=Teatro d'Annunzio;
		ch=Chieti;
		fr=Francavilla al Mare}

\end{table}

\begin{figure*}
	\centering
	{\includegraphics[width=8cm, height=7cm]{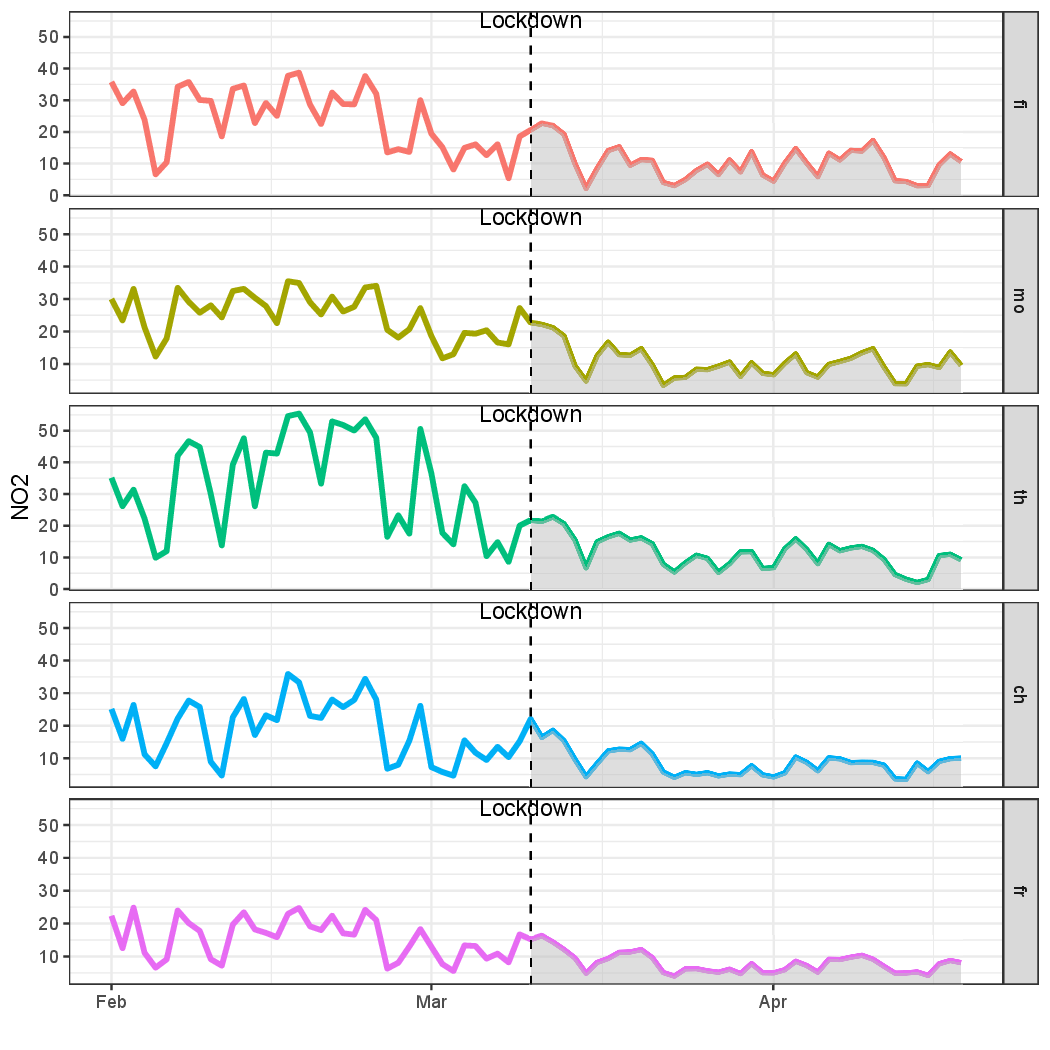}}
	{\includegraphics[width=8cm, height=7cm]{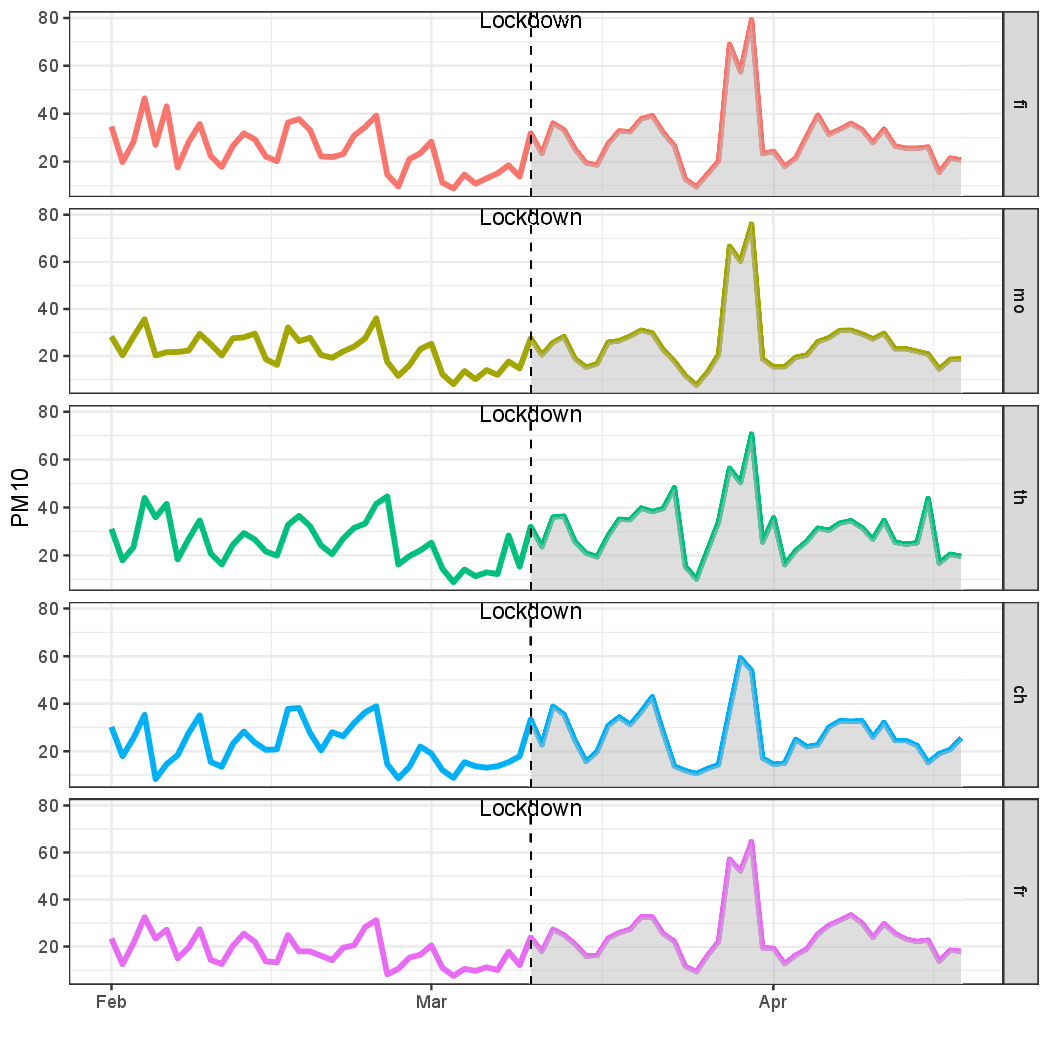}} \\
	{\includegraphics[width=8cm, height=7cm]{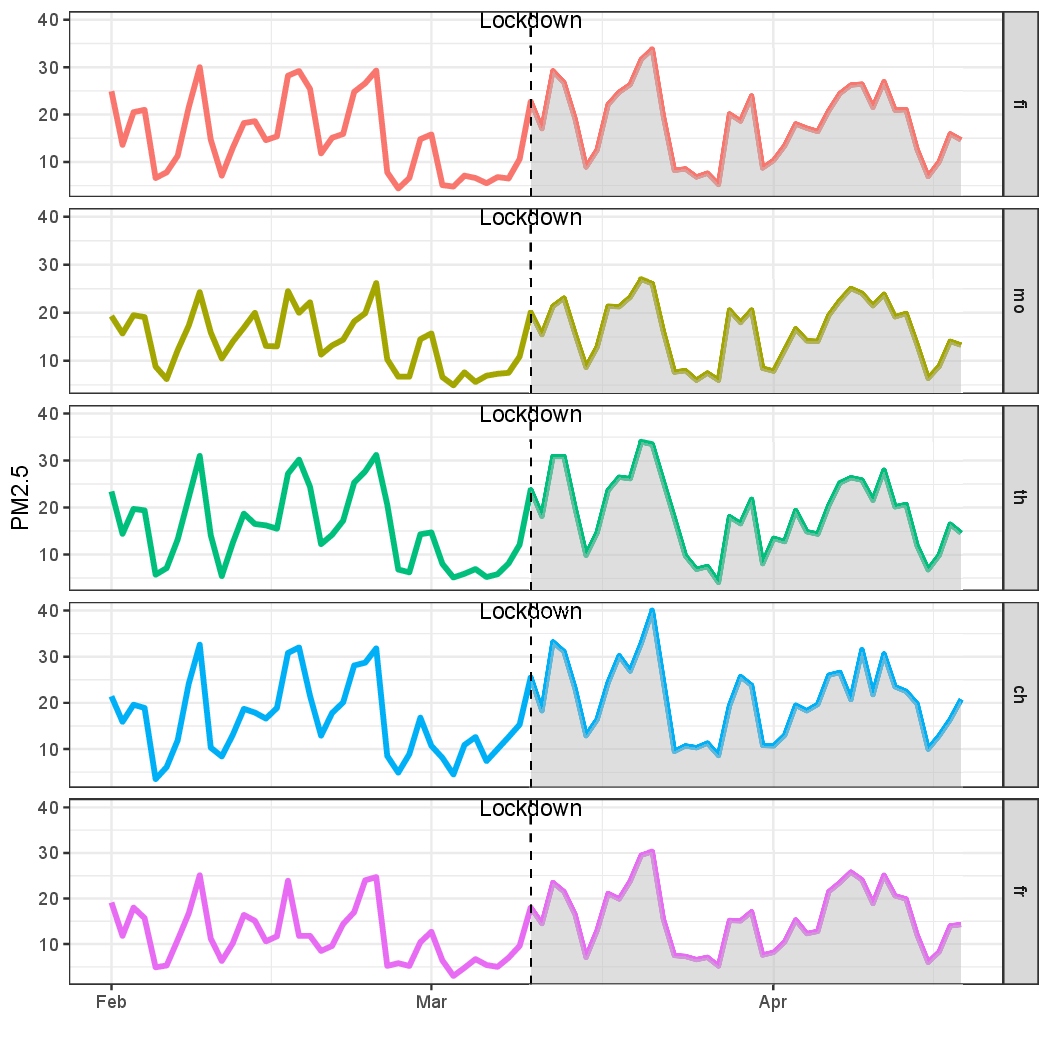}}%
	{\includegraphics[width=8cm, height=7cm]{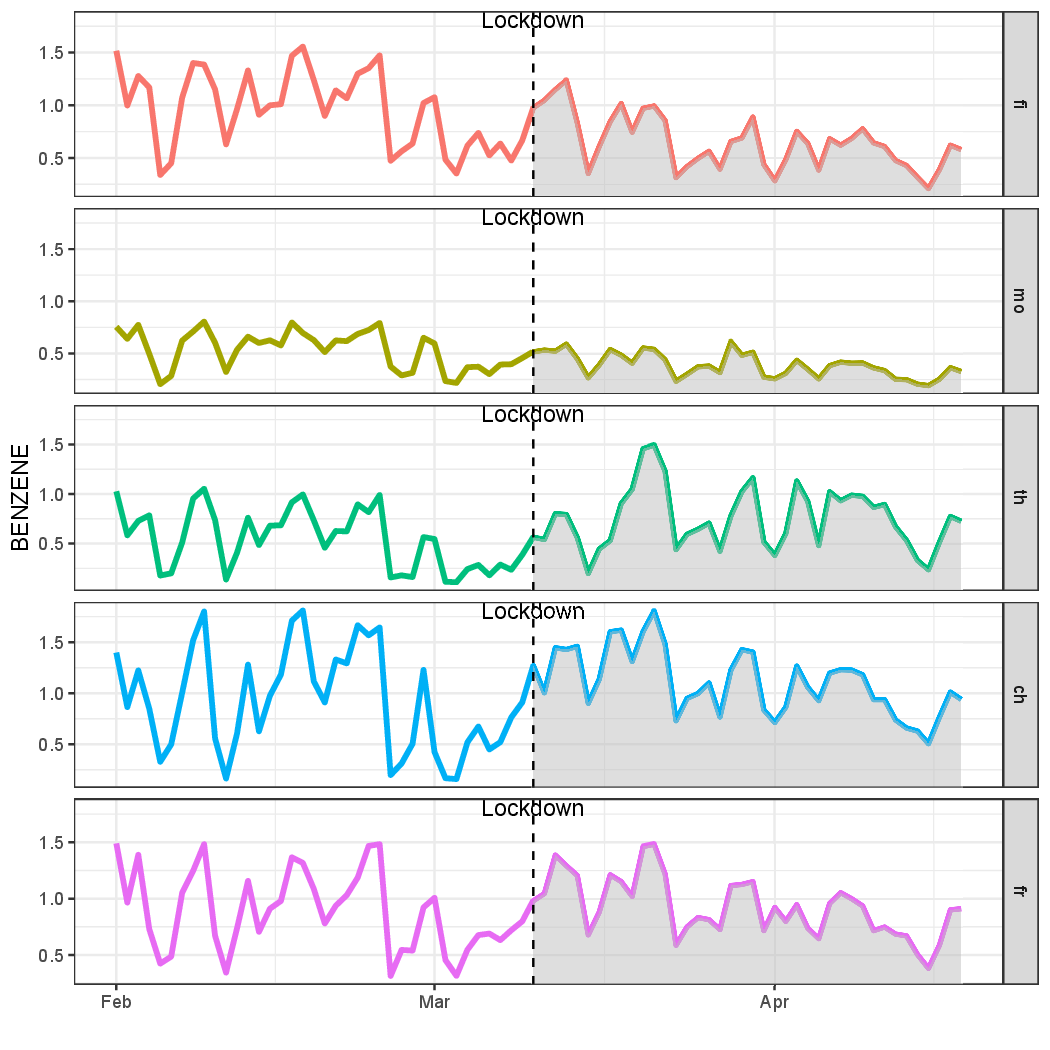}}%
	\caption{Daily variation of pollutants for stations before and during lockdown occurred on 10th March}
	\label{fig:preduring}
\end{figure*}

\begin{figure}
	\small
	{\includegraphics[width=8cm, height=7cm]{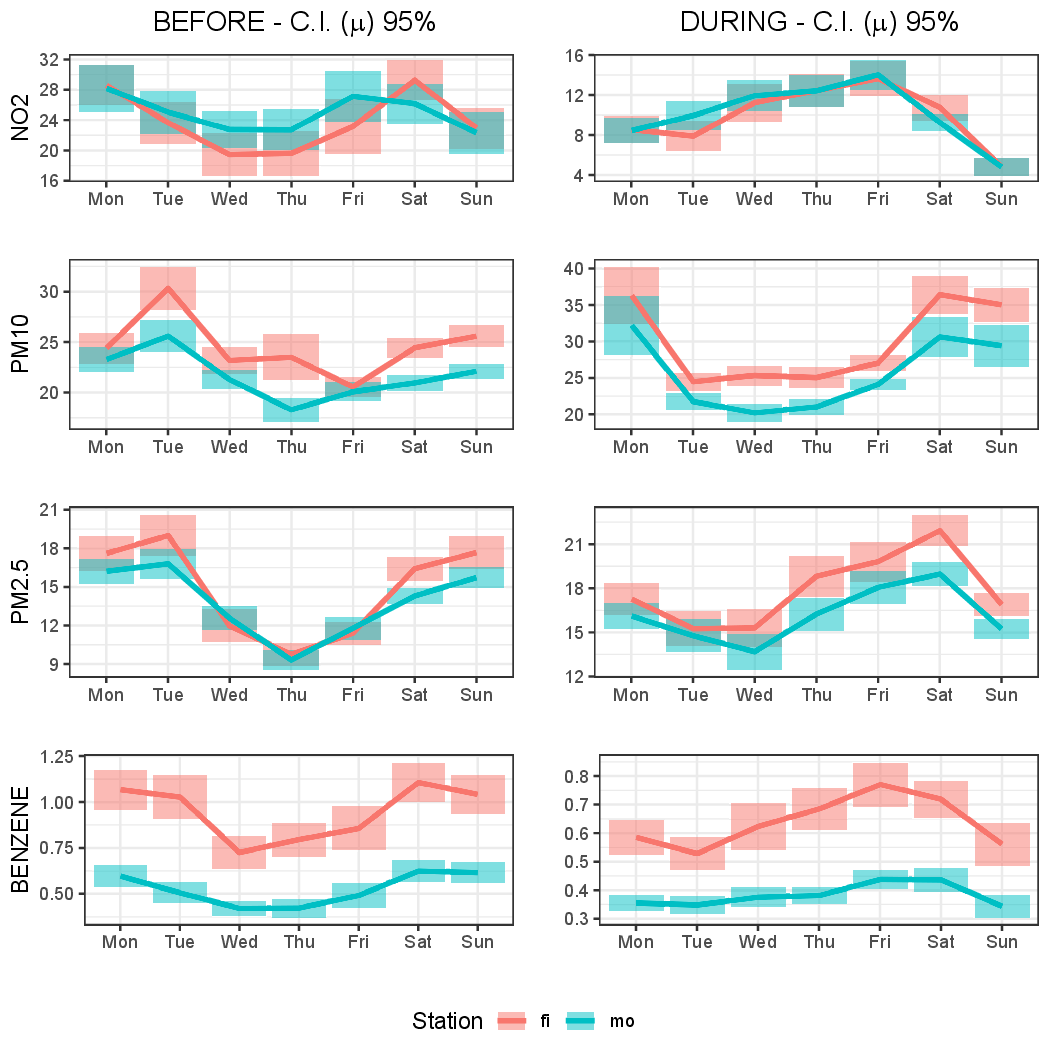}}%
	\caption{Weekly concentrations of each pollutant at traffic stations before (left panel) and during (right panel) lockdown}
	\label{fig:weekt}
\end{figure}

\begin{figure}
	\small
	
	{\includegraphics[width=8cm, height=7cm]{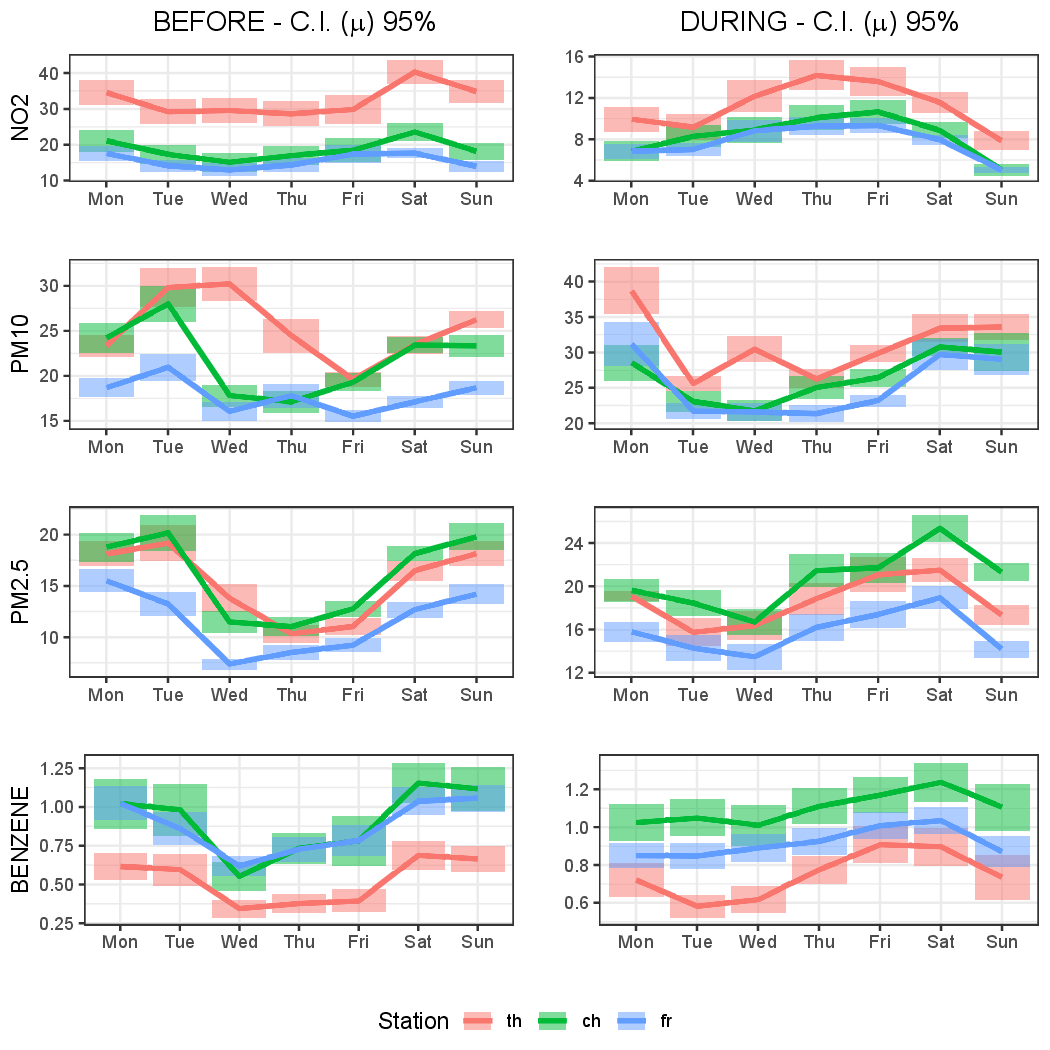}}%
	\caption{Weekly concentrations of each pollutant at background stations before (left panel) and during (right panel) lockdown}
	\label{fig:weekb}
\end{figure}

Figure \ref{fig:preduring} allows to capture the changes in concentrations of four pollutants for the pre-lockdown
and during-lockdown period.
The reduction of NO$_{2}$ during the lockdown is clearly visible and marked in all monitoring sites and  is due to the collapse of vehicular flows,
even if differences in magnitude exist depending on the stations.
As observed by several researches in other areas worldwide \citep{Wang2020,Donzelli2020}, the particulate matter (PM$_{10}$ and PM$_{2.5}$) seems to be rather independent from the measures adopted during the COVID-19 nation-wide lockdown: our data show that
background and traffic stations undergo an increase, whereas benzene levels dropped in the UT stations and increased in UB ones.
This apparently strange behaviour of particulate matter can be explained taking into account the peculiarity of this pollutant.
Airborne particulate matter can be regarded as a complex mixture of solid particles and liquid droplets generated by a wide variety of natural or anthropogenic sources, with different size and chemical composition. The fine fraction (PM$_{2.5}$) arises mainly from combustion (primary PM$_{2.5}$) or gas-to-particle conversion processes (secondary PM$_{2.5}$), while the coarse fraction of PM$_{10}$ arises mainly from traffic -  linked mechanical processes (road dust resuspension, brake and tyre wear emissions) \citep{Grigo2014} or comes from natural sources (marine aerosol, wind-blown  soil, pollens).
This implies that the monitoring sites might be under the effect of multiple emission sources, not linked to urban traffic.
In this regard, we must note that in the lockdown period the average temperature in the area under study was only slightly higher than in the pre-lockdown period (11.1 $C^{\circ}$ \emph{vs} 10.8 $C^{\circ}$, measured in Pescara at ARTA weather station), despite the fact that lockdown occurred in the first part of spring, while the previous period occurred in February and the first decade of March. This fact, together with the prolonged stay of the people at home, could have raised the emissions related to domestic heating.
Besides, a pertinent amount of PM$_{10}$ and PM$_{2.5}$ variability has a meteorological origin. As known \citep{Querol2004,Galindo2011,Wang2020}, meteorological conditions play key roles in physic-chemical processes governing formation and transport of airborne pollutants and their variability may have nullified the reductions caused by the drop in emissions related to road traffic.
Considering, for example, the concentrations of PM$_{2.5}$ in the "Teatro d'Annunzio" station, we note that the average value in the month of February and the first decade of March 2019 was 23.6 ($\mu$g/m$^{3}$), while in the same period of 2020 (the "pre lockdown period" of our analysis) the average was 15.3 ($\mu$g/m$^{3}$). This is just the effect of interannual variability, due to the different meteorological scenarios.
As an extreme example of variability induced by weather conditions, we must mention a notable long-distance transport event of natural dust from the Central Asia (Aral Sea) occurred at the end of March (29-31 March), which produced a sudden increase of the PM$_{10}$ level (clearly visible in the graphs of Figure \ref{fig:preduring}).
Concerning  the benzene concentrations, as stated earlier, it appears that this pollutant exhibits very different behaviours in background stations compared to traffic ones. In the latter, there is a decline, albeit contained, due to the reduction of vehicular flows during the lockdown, while in the former there is stability or even slight increases, probably due to domestic heating systems, particularly those fuelled by wood or pellets.
To get a comprehensive understanding of how lockdown policies have affected air pollution, we also look at the weekly concentrations of each pollutant at background and traffic stations before and during the restriction periods.
From Figures \ref{fig:weekt} and  \ref{fig:weekb}, it is evident that on Sunday, the traffic during the lockdown phase is virtually zero, therefore the concentrations of NO$_{2}$, pollutant specifically linked to vehicle emissions, reduce more than in the other days of the week.
It is worth noting that "Teatro d'Annunzio" measuring station seems to be affected by traffic emissions in an anomalous way (background site).
Also the inspection of weekly concentrations reveals that the impact of restrictions measures on PM$_{10}$ and PM$_{2.5}$ is the most complex of the four pollutants studied: we are not able to detect consistent patterns with vehicular flows even if the small sample size, only five weeks and half, could  affect the empirical findings.
More in detail, among the background monitoring stations, "Teatro d'Annunzio" results more subject to the natural component of PM$_{10}$, probably due to marine aerosol:
this station is about 200 m from the sea, with no buildings in the way.
Regarding the weekly benzene concentrations, the comparison between the two traffic monitoring sites indicates that "Via Firenze" is probably
subject to the emissions arising from combustion (mainly domestic heating) compared to those due to traffic road.
The results of FANOVA (see subsection \ref{sec:multifanres}) seem to point out a possible misclassification of this monitoring site, which is supposed to be more prone to urban traffic.
Conversely, the traffic station "Montesilvano" being on the edge of the urban area is especially exposed to road traffic emissions.

\section{Theoretical framework}\label{sec:thframe}

  In functional data analysis, the data are curves or more general functions that evolve over time, space or other continuous argument. However, the available data are usually vectors associated with the observation of a variable at a finite set of time points  (longitudinal data). In fact, functional data can be seen as a particular case of high-dimensional data with a number of highly correlated variables that is usually much larger than the sample size. Because of this, the classical multivariate  statistical methods are not usually efficient for functional  data on account of problems related to the sample size and overfitting. In particular, it is well known that the multivariate parametric (ANOVA) and nonparametric approaches for testing  homogeneity of both, independent and non independent (repeated measures) samples of vectors, do not work properly when the number of variables is large and many of them are not applicable when the dimension of the data exceeds the sample size (see, for example, \citet{Biswas2014}). The functional data analysis approaches solve these problems by introducing  in the statistical models the true form of curves and considering the metric associated with the functional space they belong to. A detailed comparison of eleven existing functional tests for the one-way ANOVA problem for independent functional data was developed in an exhaustive simulation study from which guidelines for using  the tests in practice were provided \citep{Gorecki2015}. Different FANOVA approaches are proposed in this section  for univariate   and multivariate samples of dependent and independent curves, respectively.

Let $\boldsymbol{X}_{ijr}=(X_{ijr1},...,X_{ijrH}), i=1,...,g, j=1,...,n_i$, $r=1,...,R$ be a sample of curves. Note that $g$ represents the number of independent groups, $H$ is the
number of observed response variables, $R$ denotes the number of different periods of time (or conditions) where the response variable is observed (repeated measures) and $n=\sum_{i=1}^g n_i$ is the sample size. It is considered that these curves are realizations of a $H$-dimensional stochastic process $\boldsymbol{X}=(X_1, ..., X_H)$, whose components are second order and continuous in quadratic mean stochastic processes with sample paths belonging to the Hilbert space $\mathrm{L}^2[\mathcal{T}]$ of squared integrable functions on $\mathcal{T}$, with the natural inner product
$$
<f,g>=\int_\mathcal{T}f(t)g(t)dt \ , \ \forall f,g\in\mathrm{L}^2[\mathcal{T}].
$$

\subsection{FANOVA for repeated measures} \label{sec:multifanrep}
The goal is to test the equality of mean functions associated with the observation of a functional variable  in two different conditions or periods of time for the same subjects. For instance, the problem laid out in this paper about the evolution of the quality of the air before and during the lockdown. That is, whether the level of each pollutant has changed during the lockdown. The theoretical framework involves the use of tools for repeated measures, and in particular, the analysis of variance.  In the literature there are not many works related to this matter for the field of FDA. \citet{MartinezCamblor2011} introduced the first testing procedure for this problem by keeping in mind the between group variability.  They proposed three different approaches in order to approximate the null distribution. The first technique consisted of applying a bootstrap parametric method through re-sampling some Gaussian process involved. The second and third methods were based on non-parametric approaches via bootstrap and permutation tests.  Later, \citet{Smaga2019} proposed another perspective focused on the Box-Type approximation.  In that study, the four existent methods were compared, turning out to be the Box-Type approximation the quickest option from the computational viewpoint. In relation to size control and power all of them gave similar results, and its finite sample behaviour was very satisfactory. However, both works agree that for very small sample size, the bootstrap tests are lightly nonconservative. \citet{2020} adapted two new statistics from the classical paired $t$-test to the functional data framework. This new approach is more powerful than the testing procedures aforementioned because takes the within group variability into account as well. Here, the distributions of the statistics were also approximated by parametric methods based on derived asymptotic distributions as well as non-parametric bootstrap and permutation approaches.  The simulation study proved that the asymptotic and Box-Type tests are not recommended because of their liberality. \citet{2020} suggested the permutation tests, although the non-parametric bootstrap methods also work correctly. Nevertheless, it was emphasized that there are evidences that the procedures proposed tend to be nonconservative for small sample size.

In what follows, a single functional variable is considered because they are going to  be dealt separately. In this context, it is assumed that the sample functions can be represented as $X_{jr}(t)$ with $t\in \mathcal{T}=[a,b]$, $j=1,...,n$ and $r=1,...,R$, such that $E[X_{jr}(t)]=\mu_r(t)$. Only two different conditions or periods of time are evaluated in the current work ($R=2$). Besides, each trajectory can be expressed as $X_{jr}(t)=\mu_r(t)+e_{jr}(t)$ where $e_{jr}(t)$ are random functions centered in mean. In this kind of problem the pursued goal is to test the hypothesis
$$
\left\lbrace
\begin{array}{l}
H_0: \mu_1(t)=\mu_2(t) \ \forall t\in [a,b] \\
H_1: \mu_1(t)\neq \mu_2(t) \ for \ some \ t .\\
\end{array}
\right.
$$

\citet{MartinezCamblor2011} proposed the following statistics in order to solve the statistical hypothesis testing
$$
\mathcal{C}_n=n\int_T (\overline{X}_1(t)-\overline{X}_2(t))^2 dt ,
$$

\noindent where $\overline{X}_r(t)=n^{-1}\sum_{j=1}^n X_{jr}(t)$ is the mean function for each condition or period of time. This statistics avoid the homoscedasticity assumption.

Due to $\mathcal{C}_n$ only takes the between group variability, \citet{2020} proposed the following two statistics in order to consider both the between and within group variabilities
$$
\mathcal{D}_n=n\int_T \frac{\left(\overline{X}_1(t)-\overline{X}_2(t)\right)^2}{\hat{K}(t,t)} \ dt ,
$$

$$
\mathcal{E}_n=sup_{t\in [a,b]}\left\lbrace \frac{n\left(\overline{X}_1(t)-\overline{X}_2(t)\right)^2}{\hat{K}(t,t)} \right\rbrace ,
$$

\noindent with $\hat{K}(t,t)=\frac{\sum_{j=1}^n\left[(X_{j1}(t)-\overline{X}_1(t))-(X_{j2}(t)-\overline{X}_2(t))\right]^2}{n-1}.$

One of the biggest problems in the practice is that curves are observed in discrete time because it is impossible to observe a set of functions continuously in time. Thus, the first step would be to reconstruct the functional form of the curves approximately. \citet{Ferraty2006} proposed to use non-parametric techniques for this purpose, meanwhile \citet{Ramsay2002, Ramsay2005} suggested an approach based on basis expansion of each sample curve. This last strategy consists of assuming that curves belong to a finite-dimension space spanned by a basis $\lbrace \phi_1(t),...,\phi_p(t)\rbrace$, so that they can be expressed as
$$
X_{jr}(t)=\sum_{k=1}^pa_{jrk}\phi_k(t)=\boldsymbol{\mathrm{a}}_{jr}'\boldsymbol{\mathrm{\phi}}(t)\ , \ j=1,...,n; r=1,2,
$$
\noindent where $a_{jrk}$ represent the basis coefficients of the reconstruction for the corresponding sample curve with $\boldsymbol{\mathrm{a}}_{jr}=(a_{jr1},...,a_{jrp})'$ and $\boldsymbol{\mathrm{\phi}}(t)=(\phi_1(t),...,\phi_p(t))'$. Note that $p$ must be sufficiently large to guarantee an accurate precision. Besides, it is necessary to choose properly the dimension and the type of the basis by keeping in mind the nature of the curves. There are numerous basis systems but the most employed ones are Fourier functions (for periodic data), B-spline (for non-periodic and smooth data) and wavelets (for curves with strong local behaviour). Finally, sample trajectories can be observed with error or without error. For the first case, least squares approximation is usually used in order to estimate the basis coefficients, whereas for the second scene some interpolation method could be applied. For more details about these methodologies, \citet{Ramsay2005} carried through an exhaustive study and \citet{Ramsay2009} extended theses aspects to the software R.

$\mathcal{C}_n$, $\mathcal{D}_n$ and $\mathcal{E}_n$ can be computed by considering the basis expansion. In fact, it is direct to prove that
$$
\left(\overline{X}_1(t)-\overline{X}_2(t)\right)^2=\left( \overline{\boldsymbol{\mathrm{a}}}_1'\boldsymbol{\mathrm{\phi}}(t) -\overline{\boldsymbol{\mathrm{a}}}_2'\boldsymbol{\mathrm{\phi}}(t) \right)^2=
$$
$$
\left(\boldsymbol{\mathrm{\phi}}(t)'\overline{\boldsymbol{\mathrm{d}}} \right)^2=\boldsymbol{\mathrm{\phi}}(t)'\overline{\boldsymbol{\mathrm{d}}}\overline{\boldsymbol{\mathrm{d}}}'\boldsymbol{\mathrm{\phi}}(t),
$$

\noindent and
$$
\hat{K}(t,t)=\frac{\sum_{j=1}^n\left[(X_{j1}(t)-\overline{X}_1(t))-(X_{j2}(t)-\overline{X}_2(t))\right]^2}{n-1}
$$
$$
=Var(X_1(t))-2Cov(X_1(t),X_2(t))+Var(X_2(t))
$$
$$
=\hat{C}_1(t,t)-2\hat{C}_{12}(t,t)+\hat{C}_2(t,t)
$$
$$
=\boldsymbol{\mathrm{\phi}}(t)'(\hat{\Sigma}_1-2\hat{\Sigma}_{12}+\hat{\Sigma}_2)\boldsymbol{\mathrm{\phi}}(t),
$$

\noindent with $\overline{\boldsymbol{\mathrm{d}}}=(\overline{d}_1,...,\overline{d}_p)'=\overline{\boldsymbol{\mathrm{a}}}_1-\overline{\boldsymbol{\mathrm{a}}}_2=(\overline{a}_{11},...,\overline{a}_{1p})'-(\overline{a}_{21},...,\overline{a}_{2p})'$ where $\overline{a}_{rk}=n^{-1}\sum_{j=1}^n a_{jrk}$ $r=1,2$; $k=1,...,p$. Besides, $\hat{\Sigma}_r$ is the sample covariance matriz of the matrix $A_{r}$ of basis coefficients in the group $r$, whose elements are $A_r=(a_{jrk})$, and $\hat{\Sigma}_{12}$ is the sample cross-covariance matrix between $A_1$ and $A_2$. Note for major clarity that $\overline{X}_r=n^{-1}\sum_{j=1}^n \boldsymbol{\mathrm{a}}_{jr}'\boldsymbol{\phi}(t)=\overline{\boldsymbol{\mathrm{a}}}_r'\boldsymbol{\phi}(t)$.

\subsection{Multivariate FANOVA for independent measures} \label{sec:fanrep}
The following idea in this kind of analysis is a little bit different to the case of repeated measures.  Now, the aim is to test the the equality of  the mean functions coming from independent groups. For example, the evolution of level of benzene in the air in two different regions. If there is a response variable (e.g. level of benzene), the problem is known as Univariate FANOVA. Likewise, another fundamental aspect in these studies is the number of factors that determine the different groups. If it only exists one factor (e.g. regions) the problem is called One-Way FANOVA. There are several existing methods for testing the one-way FANOVA problem \citep{Faraway1997, Cuevas2004, Zhang2013,Zhang2014}. On the other hand, \citet{Gorecki2015} made a detailed comparison of tests for the one-Way ANOVA problem for functional data and presented tests based on a basis function representation. These tests were inspired by the idea of the B-Spline method of \citet{Shen2004}. In this line, \citet{Aguilera2021}
suggested a novel approach by using Functional Principal Component Analysis (FPCA). This method consists of testing multivariate homogeneity on a vector of principal components scores. However, although there are available many works for the univariate case, the natural extension for the multivariate case (more than one functional response variable) is not a theme that it had been studied deeply. Some references are \citet{Jacques2014}, and \citet{Gorecki2017}. Here, a novel approach based on multivariate FPCA is considered for dealing with the multivariate FANOVA problem. This new methodology can be seen as the extension of the parametric and nonparametric approaches proposed by \citet{Aguilera2021}.

Suppose that $X_{ijh}(t)$ with $i=1,...,g$, $j=1,...,n_i$ and $h=1,...,H$ are a set of curves. Then, the information for each subject is a vector of curves denoted by $\boldsymbol{X}_{ij}(t)=(X_{ij1}(t),...,X_{ijH}(t))'$. Besides, it is assumed that $\boldsymbol{X}_{ij}(t)$ are i.i.d. multivariate functional variables  with mean vector  $\boldsymbol{\mu}_i=(\mu_{i1}(t),...,\mu_{iH}(t))'$ and matrix covariance function  $\boldsymbol{\mathrm{C}}$ such that $\boldsymbol{\mathrm{C}}(t,s)=(C_{h,h'}(t,s)),$ $t,s\in \mathcal{T}$ and $h,h'=1,...,H$. Note that if $h=h'$, then $C_{h,h}$ is the covariance function and otherwise, that is $h\neq h'$, $C_{h,h'}$ represents the cross-covariance function. Now, the aim is to test
$$
H_0: \boldsymbol{\mu}_1(t)=...=\boldsymbol{\mu}_g(t) \ \forall t\in [a,b],
$$

\noindent against the alternative that its negation holds.

In the field of FDA, it is very common to deal with high dimension data. These type of data are defined as data associated to a great number of highly correlated variables where the sample size is too much small. For this reason, one of the most important technique in FDA is FPCA. This tool reduces the dimension of the problem and explains the main characteristics and modes of variation of the curves in terms of a reduce set of uncorrelated variables call functional principal components (PC's).
\citet{Ramsay2002} presented the univariate approach and discussed the extension of FPCA to the case of bivariate functional data. This theory can be adapted for more than two response variables. PC's are obtained as some generalized linear combinations of the process variables with maximum variance. Formally, the $m$-th principal component scores are determined by
$$
\xi_{ijm}=\int_\mathcal{T}(\boldsymbol{X}_{ij}(t)-\boldsymbol{\mu}(t))'\boldsymbol{f}_m(t) dt=
$$
$$
\sum_{h=1}^H\int_\mathcal{T}(X_{ijh}(t)-\mu_h(t))f_{mh}(t)dt,
$$
\noindent where $\boldsymbol{\mu}(t)=(\mu_1(t),...,\mu_H(t))$ is the overall mean function and $\boldsymbol{f}_m(t)=(f_{m1}(t),...,f_{mH}(t))'$ are the vector of weight functions (or loadings) that maximizes the variance subject to $\sum_{h=1}^H \int_\mathcal{T}f_{mh}(t)f_{m'h}(t)dt=1$ if $m=m'$ and 0 otherwise. These functions are obtained as the eigenfunctions of the eigenequation system $C\boldsymbol{f}_m=\lambda_m\boldsymbol{f}_m$, with $C$ being the covariance operator and the sequence $\lbrace \lambda_m\rbrace_{m\geq1}$ of positive real eigenvalues decreasing to zero indicate the amount of variance attributable to each component. The aforementioned system can be written in detail as follows:
$$
\scriptsize{
\begin{array}{c}
\int_\mathcal{T}C_{11}(s,t)f_{m1}(t)dt+... +\int_\mathcal{T}C_{1H}(s,t)f_{mH}(t)dt=\lambda_m f_{m1}(s) \\
\int_\mathcal{T}C_{21}(s,t)f_{m1}(t)dt+... +\int_\mathcal{T}C_{2H}(s,t)f_{mH}(t)dt=\lambda_m f_{m2}(s) \\
\vdots\\
\int_\mathcal{T}C_{H1}(s,t)f_{m1}(t)dt+... +\int_\mathcal{T}C_{HH}(s,t)f_{mH}(t)dt=\lambda_m f_{mH}(s) .
\end{array}
}
$$
Highlight that each PC is a zero-mean random variable with maximum variance and uncorrelated with the remainder of PC's. Hence, in multidimensional context and similar to the univariate setting, this process admits the following orthogonal decomposition known as Karhunen-Lo\`eve expansion
$$
\boldsymbol{X}_{ij}(t)=\boldsymbol{\mu}(t)+\sum_{m=1}^\infty\xi_{ijm}\boldsymbol{f}_m(t).
$$
The principal advantage of this decomposition is that curves can be approximated by means of a principal reconstruction in terms of the first $q$ PC's, that is $\boldsymbol{X}^q_{ij}(t)=\boldsymbol{\mu}(t)+\sum_{m=1}^q\xi_{ijm}\boldsymbol{f}_m(t)$. Normally, $q$ is chosen so that the explained cumulative variability is as close as possible to 100\%.  With this approach, the dimension of the problem is considerably reduced.

\citet{Jacques2014} explain the multivariate FPCA by means of the basis expansions and it is summarized in \citet{Schmutz2020}. This approach is briefly explained hereafter. If the basis expansion is considered, $\boldsymbol{X}_{ij}(t)$ can be expressed as
$$
\boldsymbol{X}_{ij}(t)=\boldsymbol{\Phi}(t)\boldsymbol{\mathrm{a}}_{ij}',
$$
\noindent where the basis coefficients are gathered as  $\boldsymbol{\mathrm{a}}_{ij}=(a_{ij11},$ $...,a_{ij1p_1},a_{ij21},...,a_{ij2p_2},...,a_{ijH1},...,a_{ijHp_H})$ with $p_h$ being the number of basis functions for the $h$-th response variable and
$$
\tiny{\boldsymbol{\Phi}(t)=\left(
\begin{matrix}
\phi_{11}(t) & \cdots & \phi_{1p_1}(t) & 0 & \cdots & 0 & \cdots & 0 & \cdots & 0\\
0 & \cdots & 0 & \phi_{21}(t) & \cdots & \phi_{2p_2}(t) & \cdots & 0 & \cdots & 0\\
\vdots & & \vdots & \vdots & & \vdots  & \vdots & \vdots &  & \vdots \\
0 & \cdots & 0 & 0 & \cdots & 0 & \cdots & \phi_{H1}(t) & \cdots & \phi_{Hp_H}(t)\\
\end{matrix}
\right).}
$$

In general $\boldsymbol{X}(t)=\boldsymbol{A}\boldsymbol{\Phi}'(t)$, where $\boldsymbol{A}$ is the resultant matrix after joining by row all $\boldsymbol{\mathrm{a}}_{ij}$. Thus,  whether the mean vector is subtracted to each row of $\boldsymbol{X}(t)$, the spectral decomposition of the covariance operator $C$ becomes
$$
\boldsymbol{\Phi}(s)\Sigma_A\boldsymbol{W}\boldsymbol{b}_m'=\lambda_m\boldsymbol{\Phi}(s)\boldsymbol{b}_m',
$$
\noindent with $\Sigma_A$ being the covariance matrix of $\boldsymbol{A}$,  $\boldsymbol{b}_m$ being a row-vector that contains the basis coefficients of $\boldsymbol{f}_m(t)=\boldsymbol{\Phi}(t)\boldsymbol{b}_m'$ and $\boldsymbol{W}=\int_\mathcal{T}\boldsymbol{\Phi}(t)'\boldsymbol{\Phi}(t)dt$ being the matrix of inner products between
basis functions with dimension $\sum_{h=1}^Hp_h \times \sum_{h=1}^Hp_h$. Since the showed spectral decomposition is true for all $s$, the expression can be reduced as $\Sigma_A\boldsymbol{W}\boldsymbol{b}_m'=\lambda_m\boldsymbol{b}_m'$. Now, by considering $\boldsymbol{u}_m=\boldsymbol{b}_m\boldsymbol{W}^{1/2}$, the multivariate FPCA is equivalent to the multivariate PCA of the matrix $\boldsymbol{A}\boldsymbol{W}^{1/2}$, whose covariance matrix can be diagonalized as follows:
$$
\boldsymbol{W}^{1/2'}\Sigma_A\boldsymbol{W}^{1/2}\boldsymbol{u}_m'=\lambda_m\boldsymbol{u}_m'.
$$
Therefore, the PC's are given by
$$
\xi_{ijm}=\boldsymbol{\mathrm{a}}_{ij}'\boldsymbol{W}\boldsymbol{b}_m.
$$

In order to obtain the principal component scores for the multivariate case, \citet{Schmutz2019} implemented in the software R \citep{RLanguage2020} the package called "\textit{funHDDC}". Once they are computed, there are suggested two different ways to solve the problem based on testing homogeneity on the  vector of the first  $q$ principal components scores in the $g$ groups. The first one is to perform univariate ANOVA on each principal component correcting the level of significance for the normality case. It is well-known whether the multivariate normality is suitable, the uncorrelatedness implies independence and then, it does not make sense to consider a multivariate approach. Otherwise, when the multivariate normality is not satisfied, the option is to apply non-parametric multivariate tests such as the extensions of the univariate Kruskal Wallis's test and Moods's test. Normally, it is recommended to use the permutation version of the tests when the sample size is small \citep{Oja2010}.


\section{Results}\label{sec:multfan}
We now illustrate the use of testing procedures previously described to ascertain whether the level of each pollutant has changed during the lockdown
period.
As pointed out in Section \ref{sec:data}, to reveal the impact of restriction measures due to the COVID-19 on the air quality, the obtained environmental datasets were divided
in two time frames, of the same length (39 days): i) pre-lockdown (February 1, 2020 - March 10, 2020) and ii) during-lockdown (March 11, 2020 - April 18, 2020). From a theoretical viewpoint, we have longitudinal functional data corresponding with the observation of the same functional variables in two different periods of time.

\subsection{Functional reconstruction of pollutant curves}
\begin{figure}
\includegraphics[width=8cm, height=7cm]{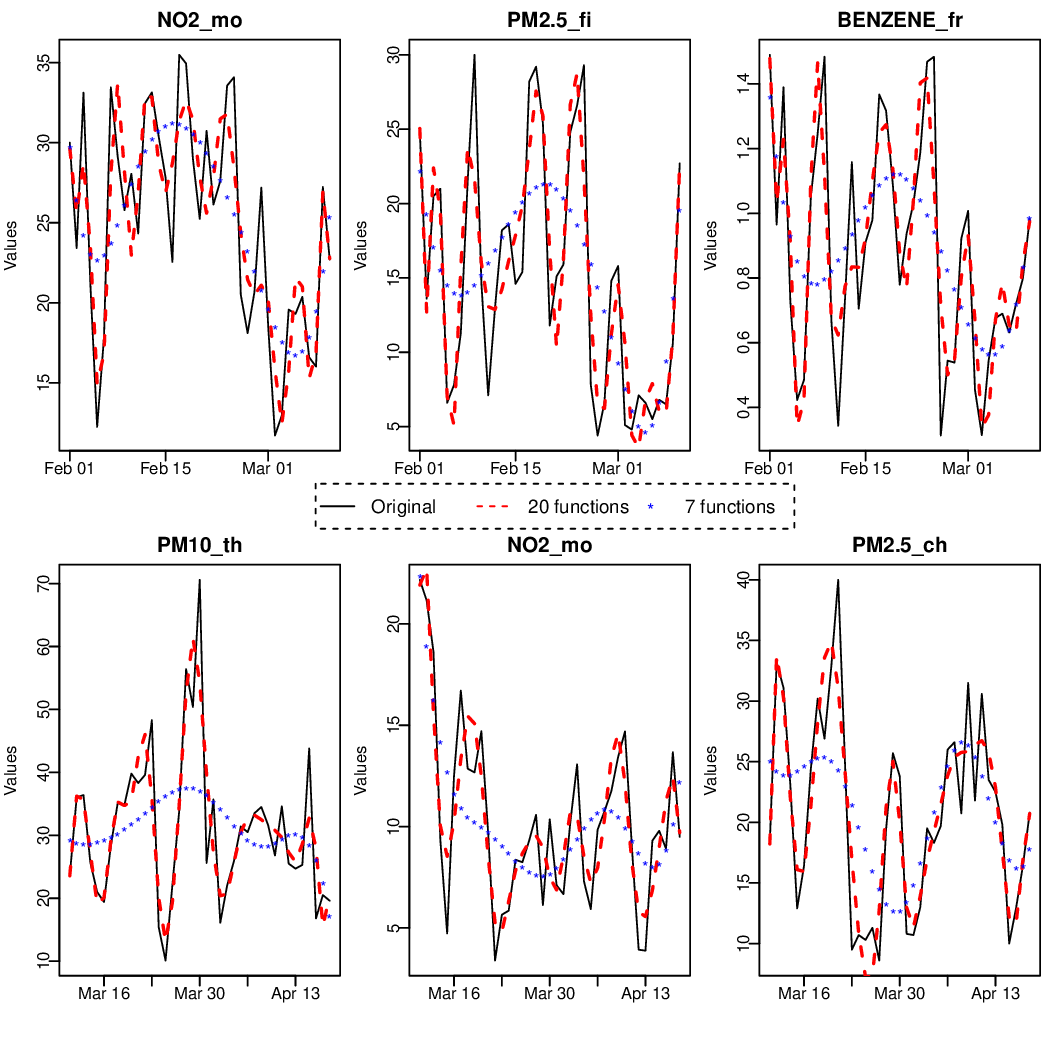}
\caption{Functional approximation, using 7 and 20 basis functions, of some pollutants for stations before lockdown (upper panel) and during lockdown (lower panel)}
\label{fig:7_20basis}
\end{figure}

\begin{figure}
\includegraphics[width=8cm, height=7cm]{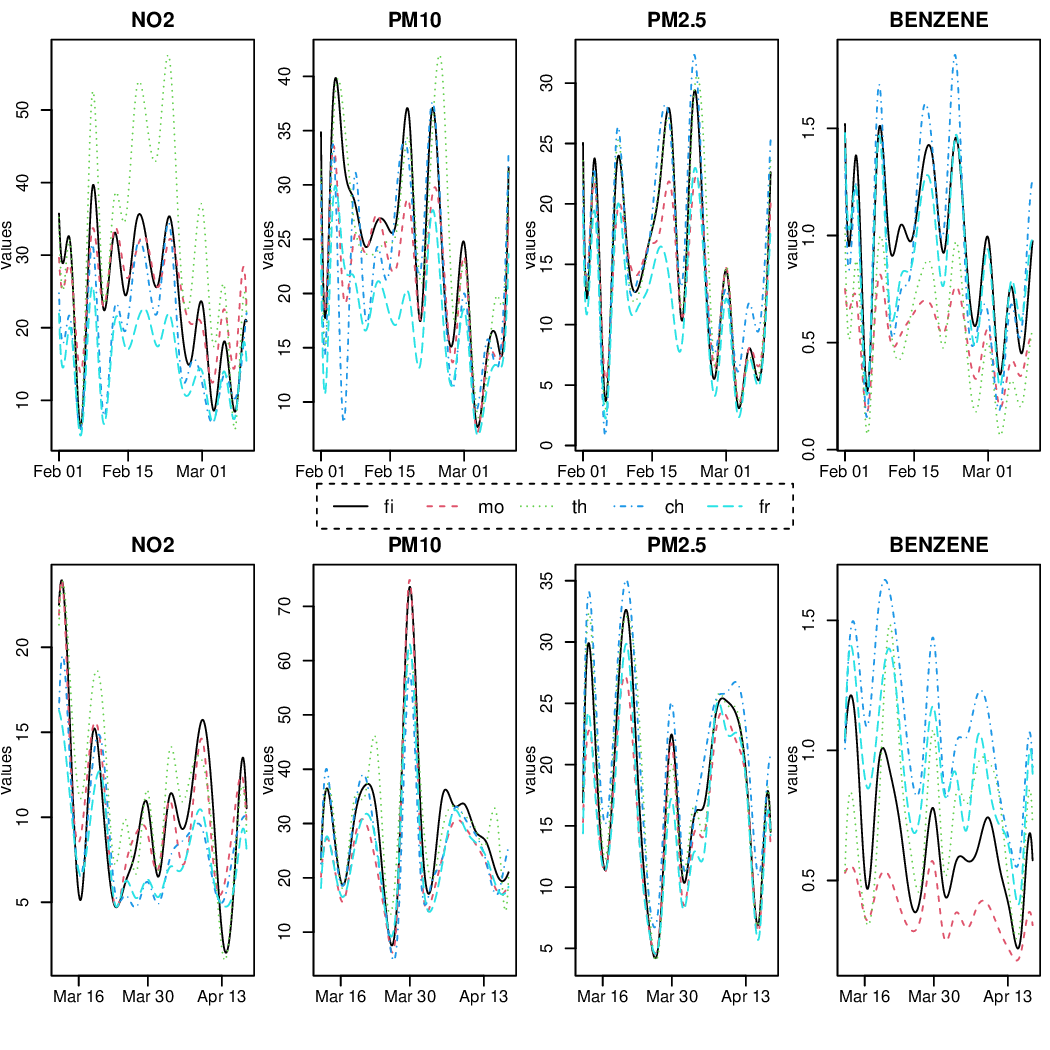}
\caption{Functional approximation, using 20 basis functions, of pollutants for stations before lockdown (upper panel) and during lockdown (lower panel)}
\label{fig:20basis_all}
\end{figure}

As a first step of our data analysis, we reconstructed the functional form of curves from the initial points that come from the discrete values measured in the study.
To convert the discretely observed data to smooth functions, the reconstruction of curves is made by using a cubic B-spline smoothing. The B-spline functions are  one of the most prominent spline basis, used for non-periodic functions, which is proven to be numerically stable and flexible \citep{Ramsay2005}.
Initially, in tailoring a basis system to fit our data, we used 7 basis functions. This option is conservative: it allows to capture the trend of curves but not their local behaviour. To recover the underlying functions of the observed data, we were increasing the number of basis functions up to 20. This choice preserves important information about the real form of the curves. Figure \ref{fig:7_20basis} illustrates the shape the data would take after smoothing them into these basis systems. It is clear that the increase of the number of basis functions produces smaller  differences between the smoothed sample curves and the observed data (see Figure \ref{fig:20basis_all}). Hereinafter, an approximation of each sample curve in terms of a basis of cubic B-splines of dimension 20 is considered. \\

\subsection{FANOVA for repeated measures results} \label{sec:fanres}

Before moving on more complex studies, we carried out a univariate analysis to evaluate the behaviour of each pollutant before and during lockdown.
To statistically confirm the effect of lockdown on the mean of each pollutant, we first implemented FANOVA for repeated measures, as defined in Section \ref{sec:fanrep}.
Specifically, we apply the statistics $\mathcal{D}_n$ and $\mathcal{E}_n$ which are the best to control the between and within group variability that there are behind the repeated measures design. In order to construct the tests based on these statistics, a permutation method is used to approximate their null distributions. This technique consists of a random permutation of each sample unit. Let us denote  the original data by  $\boldsymbol{X}=(X_{1},X_{2},\ldots,X_{n})$ where $\boldsymbol{X_j}=(X_{j,1},X_{j,2})$ $(j=1,\dots,n),$ and the resampling vectors by  $\boldsymbol{X}_*=(\boldsymbol{X_1}*,\ldots,\boldsymbol{X_n}*)$ with $\boldsymbol{X_j}*=(X_{j,1}*,X_{j,2}*)$ being  a random permutation of the sample unit $X_j.$ This process is repeated $\Delta$ times, with $\Delta$ a number sufficiently large, so that $\mathcal{D}_n^\delta*$ and $\mathcal{E}_n^\delta*$ are calculated for each replication, being $\delta=1,\ldots,\Delta$. Later, p-values are obtained as the proportion of times that $\mathcal{D}_n^\delta*$ and $\mathcal{E}_n^\delta*$ overcome $\mathcal{D}_n$ and $\mathcal{E}_n$, respectively. Here, the p-values were obtained from 2000 replications.
The results of the proposed testing procedures are shown in Table \ref{tab:fanrep}.
The p-values of all tests are less than the significance level $\alpha=0.05$ for NO$_{2}$, PM$_{10}$ and PM$_{2.5}$. For benzene, $\mathcal{E}_n$ shows no differences between both periods, but it is very close to the limit region. Therefore, and taking into account the sample size, we have evidence to reject the null hypothesis for benzene and we state that there are also differences in the mean curves of this pollutant in the pre and during lockdown phases.
These results statistically confirm the evidences already reported in Table \ref{tab:mean} and discussed in Section \ref{sec:descriptives}.
For some pollutants (PM$_{10}$ and PM$_{2.5}$), we recorded an increase during the lockdown tenure
probably due to a greater employment of domestic heating systems and the key roles of particular meteorological conditions that govern the formation and transport of particulate matter. For benzene, we pointed out a differentiate behaviour according to the type of monitoring stations, influenced by different emission sources (traffic \emph{vs }domestic heating). Finally, the collapse of vehicular traffic during the quarantine days justifies the steep decline of NO$_{2}$ in both types of measuring sites.

\begin{table}
	\caption{FANOVA for repeated measures on the test statistics  $\mathcal{D}_n$ and $\mathcal{E}_n$ }
	\vspace{0.3cm}
	\setlength{\tabcolsep}{6pt} 
	\begin{center}	
		\begin{tabular}{ll|cc}
			&	 \emph{ p-value}	&	\textbf{$\mathcal{D}_n$}		&	\textbf{$\mathcal{E}_n$}	\\ \hline
			&	NO$_{2}$	& 0.034 & 0.035 \\
			&	PM$_{10}$ 	& 0.000 & 0.034	\\
			&   PM$_{2.5}$ 	& 0.028 & 0.030	\\
			&	Benzene	&	0.049 & 0.070 	\\
		\end{tabular}
	\end{center}	
	\label{tab:fanrep}
\end{table}

\subsection{Multivariare FANOVA results for independent measures} \label{sec:multifanres}
Once the impact of the lockdown has been studied, a further step of our data analysis has involved the assessment of equality of mean functions of individual groups.
In our context, the groups have been individuated according to the location of the monitoring sites. In more detail, our interest lies in investigating if the mean function of all the pollutants measured in the background stations is equal to that of the urban traffic ones.
The multivariate analysis of variance is carried out both before and during lockdown tenure to detect differences attributable to the government restrictions.
This comparison has been evaluated firstly globally, considering all the pollutants together, and then for each variable separately.
In Table \ref{tab:fanindep}, the results for multivariate and univariate FANOVA based on FPCA are displayed. On this matter, four principal components are chosen for both cases (multivariate and univariate analysis), since more than a 99\% of total variability is explained with four components in all situations. Besides, due to the fact that the normality is in question and the sample size is really small, the extension of the univariate Kruskal-Wallis's test with the permutation version is conducted by means of \textit{`MNM'} R package.

\begin{figure}
\includegraphics[width=8cm, height=8cm]{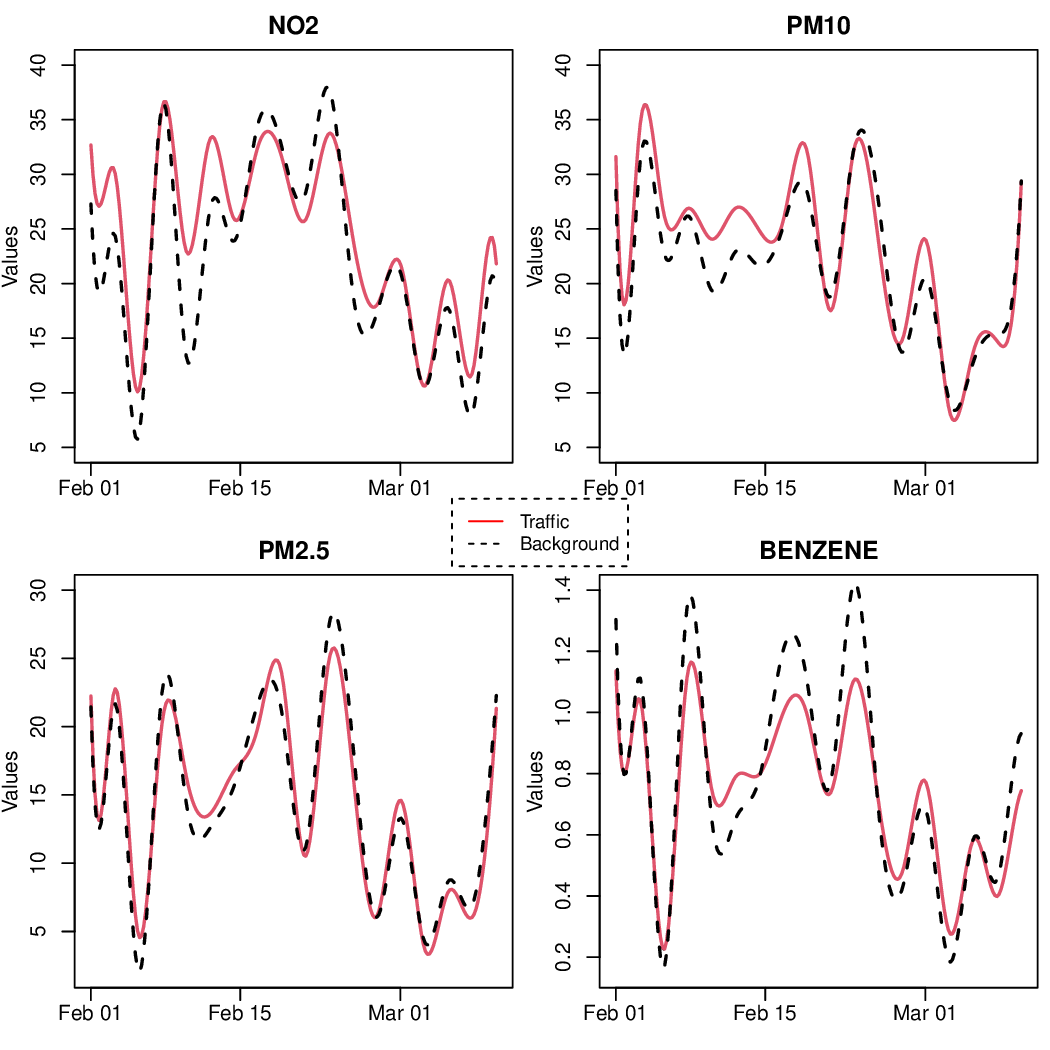}
\caption{Mean function per pollutant of each group before lockdown}
\label{fig:before_clus}
\end{figure}

\begin{figure}
\includegraphics[width=8cm, height=7cm]{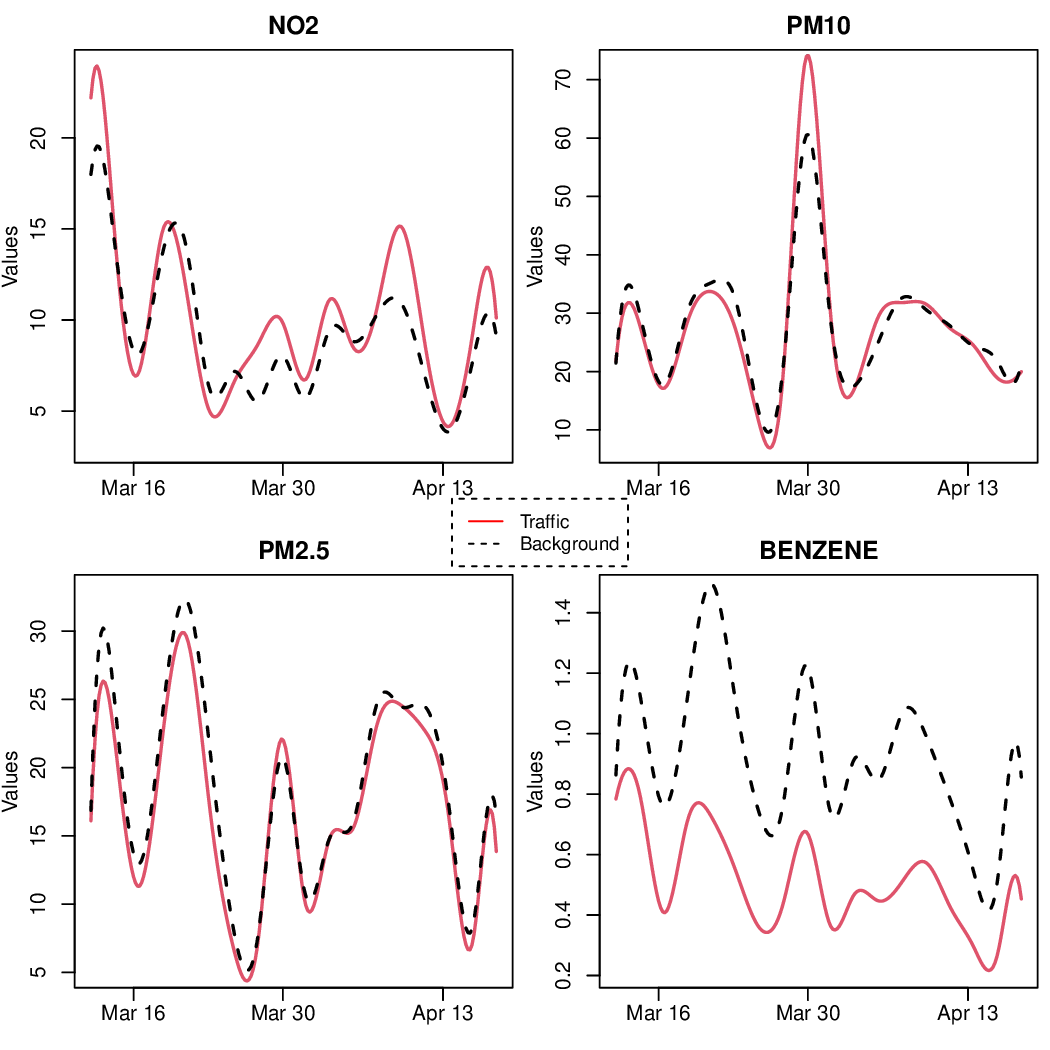}
\caption{Mean function per pollutant of each group during lockdown}
\label{fig:during_clus}
\end{figure}

\begin{table}
	\caption{Multivariate FANOVA for independent measures}
\vspace{0.3cm}
	\setlength{\tabcolsep}{6pt} 
	\begin{center}
		\begin{tabular}{ll|cc}
			&	 \emph{ p-value}	&	\textbf{BL}		&	\textbf{DL}		\\ \hline
			& All pollutants & 0.000 & 0.302 \\
			&	NO$_{2}$ & 0.562 & 0.272 \\
			&	PM$_{10}$ 	& 0.000 & 0.306 \\
			&   PM$_{2.5}$  & 0.889 & 0.685	\\
			&	Benzene	 & 0.186 & 0.000	\\
		\end{tabular}
	\end{center}
	\label{tab:fanindep}
\vspace{0.2cm}
\small{
\textbf{Acronyms:}
\vspace{0.1cm}\\
BL=Before Lockdown;
DL=During Lockdown}

\end{table}
Looking at the results of multivariate FANOVA, we found that, in the pre-lockdown phase, the groups are different from each other and the main discrimination is ascribable to the PM$_{10}$ concentrations.
Furthermore, it seems that there could be indications of significance as well regarding the benzene because the p-value is
0.186 and by increasing the sample size we could reject the homogeneity in this pollutant.
Conversely, the multivariate test is not able to distinguish the two groups in the lockdown period.
In fact, the p-value for the multivariate test is equal to 0.30. However, when we carry out the univariate tests, we record significant differences between the groups in relation to the benzene. This appreciation is also corroborated by the visual inspections of Figures \ref{fig:before_clus} and \ref{fig:during_clus}. A possible explanation for these results can be found in the simulation study performed by \citet{Aguilera2021} where it was shown that these approaches tend to be very conservative for small sample sizes.\\
As for fine particulate (PM$_{2.5}$), the results of univariate test confirm, as already known in literature, that this pollutant is ubiquitous in urban environment due to the widespread diffusion of the sources (traffic, domestic heating etc.) and the important role played by meteorological variables in the processes of its formation and transport.

\section{Conclusions}\label{sec:conclu}
Recent studies suggested that lockdown measures, adopted by the most hard-hit countries around the world in the Spring of 2020 to prevent the spread of COVID-19, have had a positive significant impact on air quality.
In this paper,  novel approaches for functional analysis of variance with univariate repeated measures and multivariate independent measures are presented, as a new methodology for a more effective understanding of the impact of lockdown on four critical air pollutants, measured in five monitoring sites in the urban area of Chieti-Pescara (Central Italy).
Being a powerful approach to modelling temporal observations, which is complementary to the usual time series techniques,  FDA allowed us to reconstruct the temporal profiles of the studied pollutants for the lockdown and unlock phases in  each measuring station.
We have found significant reduction in NO$_{2}$ levels during the lockdown period albeit some differences in magnitude are recorded according to the monitoring station.
These results are in line with the findings of other published studies on this topic
\citep{Mahato2020, Berman2020, Gautam2020,Kerimray2020}.
Unlike the NO$_{2}$ pollutant, for particulate matter, that is for PM$_{10}$ and PM$_{2.5}$, the monitoring stations experienced an increase during the quarantine weeks.
Besides, less clear was the impact of lockdown on benzene levels: the concentrations of this pollutant were smaller in the traffic stations while an increasing trend was observed
in the background measuring sites. Equally important was to determine if these differences were statistically significant.
In this respect, the functional analysis of variance has proven to be beneficial to monitoring the evolution of air quality before and during the lockdown tenure and to assessing the equality of mean functions of individual groups, individuated according to the location of measuring sites.
The considered  FANOVA approaches based on basis expansion of sample curves, dimension reduction by using FPCA of pollutants curves and testing homogeneity on the vector of the most explicative principal component scores have made this analysis feasible providing contrasted evidence to reject the null hypothesis of equality in the mean functions of all pollutants, both in the time frame considered and the localization of monitoring stations.
It should be noted that the results of multivariate FANOVA for independent measures, shown in Table \ref{tab:fanindep}, suggest a possible misclassification of the monitoring stations as concern the NO$_{2}$, because the proposed technique failed to discriminate between UB and UT measuring sites, despite the fact that NO$_{2}$, in urban areas, is a pollutant mostly produced by traffic emissions.
The acsuknowledge of the presence of some redundant or misclassified monitoring stations will provide better pport for managers to
formulate a more adequate air pollution control strategy. For all air quality monitoring networks the identification of misclassified measurements is, in fact,
an important task, not only for determining the cost
of pollution monitoring, but also for determining the
integrity of pollution information monitoring and the
accuracy of air quality assessment.
In general, the FDA framework has provided a valid understanding and knowledge of the temporal behaviour of air pollutants in a kind of controlled experiment such that offered by the lockdown.
The COVID-19 restrictions reduced the anthropogenic emissions and created an ``unprecedented scenario" in which
the source of road traffic has been drastically dropped out.\\
We believe that the results of this study are of interest for environmental protection agencies involved in developing policies to achieve air quality improvements, encouraging them to establish mechanisms to reduce pollution emissions and properly redesign local monitoring network.

\twocolumn

\section*{Aknowledgements}
This research was funded by  project
PID2020-113961GB-I00 of the Spanish Ministry of
Science and  Innovation (also supported by the FEDER program),  project FQM-307 of the Government of Andalusia (Spain)
and the PhD grant (FPU18/01779) awarded to  Christian Acal.
The authors also thank the support of the University of Granada, Spain, under project for young researchers PPJIB2020-01.

\section*{Data availability}

All data used during the study are available from the corresponding author by request.

\section*{Declarations}

\textbf{Conflict of Interest} The authors declare that they have no known competing financial interests or personal relationships that could have appeared to influence the work reported in this paper.

\section*{Abbreviations}

The following abbreviations are used in this manuscript:\\

\hspace{-0.75cm}\begin{tabular}{ll}
	
	ANOVA & Analysis of Variance \\
	
	AQI & Air Quality Index \\
	
	CO & Carbon Monoxide \\
	
	GDP & Gross Domestic Product \\
	
	FANOVA & Functional Analysis of Variance \\
	
	FDA & Functional Data Analysis \\
	
	FPCA & Functional Principal Component Analysis  \\
	
	MANOVA & Multivariate Analysis of Variance \\
	
	NMHC & Non Methane Hydrocarbons \\
	
	NH$_3$ & Ammonia \\
	
	NO$_2$ & Nitrogen Dioxide \\
	
	NO$_x$ & Nitrogen Oxide \\
	
	O$_3$ & Ground level Ozone\\
	
	PM$_{2.5}$ & Particulate Matter\\
	
	PM$_{10}$ & Particulate Matter\\
	
	PC & Principal Component\\
	
	SO$_2$ & Sulphur dioxide\\
	
	UB & Urban Background \\
	
	UT & Urban Traffic type \\
	
	VOC$_s$ & Volatile Organic Compounds\\
	
	WHO & World Health Organization\\
\end{tabular}

\bibliographystyle{plainnat}
\bibliography{Biblio_fda}

\onecolumn

\section*{Appendix}

\begin{center}
\scriptsize
\begin{longtable}{ p{3cm}p{3cm}p{4cm}p{6cm}}
 \\  \addlinespace[3pt]
\caption{A simple longtable example} \label{tab:overview}\\
\hline
\hline \multicolumn{1}{c}{\textbf{Author (year)}} & \multicolumn{1}{c}{\textbf{Study area}} & \multicolumn{1}{c}{\textbf{ Pollutant Types}} & \multicolumn{1}{c}{\textbf{Key observations}}\\ \hline
\endfirsthead
\multicolumn{4}{c}%
{\tablename\ \thetable\ -- Overview of selected studies on air pollution and COVID-19} \\
\hline \multicolumn{1}{c}{\textbf{Author (year)}} & \multicolumn{1}{c}{\textbf{Study area}} & \multicolumn{1}{c}{\textbf{ Pollutant Types}} & \multicolumn{1}{c}{\textbf{Key observations}}\\ \hline
\endfirsthead
\multicolumn{4}{c}%
{\tablename\ \thetable\ -- \textit{Overview of selected studies on air pollution and COVID-19}} \\
\hline \multicolumn{1}{c}{\textbf{Author (year)}} & \multicolumn{1}{c}{\textbf{Study area}} & \multicolumn{1}{c}{\textbf{ Pollutant Types}} & \multicolumn{1}{c}{\textbf{Key observations}}\\ \hline
\endhead
\hline \multicolumn{4}{r}{\textit{Continued on next page}} \\
\endfoot
\hline
\endlastfoot

 \\  \addlinespace[3pt]	
Lal et al. (2020) & Global & $NO_{2}$, $CO$, $AOD$ &  Reductions in $NO_2$, $CO$ were observed in the major hotspots of COVID$-$19 outbreak during Feb$-$Mar 2020. Besides, the authors paid attention to investigate the AOD level variation during COVID-19 situation. \\
 \\  \addlinespace[3pt]			
Dutheil et al.(2020) & Global and China&  $NO_{2}$ & Exploiting data from the TROPOspheric Monitoring Instrument (TROPOMI) sensor on board ESA\' s Sentinel$-$5 satellite, massive reductions in $NO_{2}$ due to quarantine were observed near Wuhan, China ($\sim$ 30\%) and worldwide. \\
 \\  \addlinespace[3pt]

Venter et al. (2020) & Global (27 countries, China, India and Europe) &$NO_{2}$,  $O_{3}$, $PM_{2.5}$  &   Using satellite data and a network of more than 10,000 air quality stations, the authors find remarkable declines in ground-level nitrogen dioxide ($NO_2$: $-$29\% with 95\% confidence interval $-$44\% to $-$13\%), Ozone ($O_{3}$: $-$11\%; $-$20\% to $-$2\%) and fine particulate matter ($PM_{2.5}$: $-$9\%; $-$28\% to 10\%) during the first two weeks of lockdown. \\
 \\  \addlinespace[3pt]
Gope et al (2021) & Global (most polluted cities worldwide) & $PM_{2.5}$ ,$PM_{10}$, $O3$, $NO_{2}$, $CO$, $SO_{2}$ & It has been detected that  the air quality of all the places has improved significantly. $NO_2$ concentration has decreased in all the cities around the world. $PM_{2.5}$ and $PM_{10}$ are the most affecting air concentration which control the air quality of all the selected places during and after lockdown. \\
 \\  \addlinespace[3pt]
Wang et al (2020) & China & $PM_{2.5}$ & Findings from this study supported that the suspension of anthropogenic activities (transportation and industry) contributed to the decrease of $PM_{2.5}$ concentrations. In this study, it is also shown that the benefits of emission reductions were overwhelmed by adverse meteorology. \\
 \\  \addlinespace[3pt]
Berman  $\&$ Ebisu (2020) & USA & $NO_{2}$, $PM_{10}$ & The changes in levels of air pollutants across USA during COVID-19 pandemic have been investigated. The authors reported a significant reduction on $NO_2$ (up to $-$25.5) and an overall decline in $PM_{2.5}$, compared with pre-lockdown phase. \\
 \\  \addlinespace[3pt]
Bao  $\&$ Zhang (2020) & 44 cities in Northern China & $SO_{2}$, $PM_{2.5}$, $PM_{10}$, $NO_{2}$, $CO$ & The vehicular restrictions during the lockdown period have led to a significant reduction of the concentrations of $SO2$, $PM_{2.5}$, $PM_{10}$, $NO_2$ and $CO$, decreased significantly by 6.76\%, 5.93\%, 13.66\%, 24.67\% and 4.58\%, respectively.\\
 \\  \addlinespace[3pt]
Li et al. (2020) & Yangtze River Delta Region (China) & $SO_{2}$, $NO_{2}$, $C0$, $0_{3}$, $PM_{2.5}$, $PM_{10}$, $VOCs$ & This study investigated the impact of reduced human activity on air quality over the Yangtze River Delta Region. During the most stringent Level I response period, primary pollutants like $SO_{2}$, $NO_2$, $PM_{2.5}$ and $VOCs$ have been reduced by 26\%, 47\%, 46\% and 57\% .\\
 \\  \addlinespace[3pt]
Zambrano-Monserrate et al. (2020) & China and Europe (France, Germany, Spain, and Italy) & $NO_{2}$, $PM_{2.5}$ & This research was aimed to highlight the positive and negative indirect effects that the new coronavirus has had on the environment. Lockdown measures led to reduced $PM_{2.5}$  and $NO_2$ concentrations. Conversely, among indirect negative effects, the increase in domestic and medical waste were mentioned. \\
 \\  \addlinespace[3pt]
Sicard et al. (2020) & Southern European cities (Nice, Rome, Valencia and Turin) and Wuhan (China) &  $NO_{x}$, $NO_{2}$, $PM_{2.5}$, $PM_{10}$, $O_{3}$ & In comparison to 2017$-$19, the lockdown measures led to a decrease of $N0_2$ ($\sim$ 53\% in Europe and 57\% in Wuhan), $NO$ ($\sim$63\% in Europe), and $PM_{2.5}$ and $PM_{10}$ ($\sim$8\% in Europe and $\sim$42\% in Wuhan) at urban stations. $NO_2$ and $NO$ decreased by $\sim$65\% and $\sim$78\% respectively at traffic stations in Europe. Conversely, $O_{3}$ increased (24\% in Nice, 14\% in Rome, 27\% in Turin, 2.4\% in Valencia and 36\% in Wuhan).\\
 \\  \addlinespace[3pt]
Collivignarelli et (2020) & Milan (Italy)& $PM_{10}$, $PM_{2.5}$, $BC$, benzene,  $CO$, $SO_{2}$, $NO_{2}$,  $NO_{x}$, $O_{3}$, $NH_{3}$ & The lockdown led to significant reduction of pollutant concentrations mainly due to vehicular traffic ($PM_{10}$, $PM_{2.5}$, $BC$, benzene, $CO$ and $N0_{x}$).\\
 \\  \addlinespace[3pt]
Tobias et al. (2020) & Spain (Barcelona) & BC, $PM_{10}$, $NO_{2}$,  $O_{3}$ & This study investigated the  changes in air pollution levels during the lockdown in terms of urban background and traffic air quality observed stations. After two weeks of lockdown, the most significant reduction was estimated for BC and $NO_2$ ($-$45 to $-$51\%), pollutants mainly related to traffic emissions. A lower reduction was observed for $PM_{10}$ ($-$28 to $-$31.0\%). By contrast, $O_{3}$, levels increased (from $+$33 to $+$57\%).  \\
 \\  \addlinespace[3pt]
Otmani et al. (2020) & Sale City (Morocco) & $PM_{10}$, $SO_{2}$, $NO_2$  & Analysing  air pollutants before and during the lockdown period, in this study it was found that $PM_{10}$, $SO_{2}$ and $NO_2$ concentrations were reduced respectively by 75\%, 49\% and 96\%. \\
 \\  \addlinespace[3pt]
Sharma  et al (2020) & India & $PM_{10}$, $PM_{2.5}$, CO,$NO_2$, $O_{3}$, $SO_{2}$ & During March $16^{th}$ to April $14^{th}$ from 2017 to 2020 in 22 cities covering different regions of India were analyzed. Overall, around 43\% , 31\% , 10\% and 18\% decreases in $PM_{2.5}$, $PM_{10}$, CO, and $NO_2$ in India were observed during lockdown period compared to previous years. \\
 \\  \addlinespace[3pt]
Mahato et al. (2020) & India (Delhi) & $PM_{10}$, $PM_{2.5}$, $SO_{2}$, $NO_{2}$, $CO$, $O_{3}$ and $NH_{3}$ & Data on seven pollutants, collected over 34 monitoring stations in Delhi, were analyzed during pre-lockdown periods and during the lockdown. Empirical findings gave evidence that air quality significantly improved during lockdown, with reductions of 60\% ($PM_{10}$), 39\% ($PM_{2.5}$), 53\% ($NO_2$) and 30\% ($CO$) compared to 2019.\\
 \\  \addlinespace[3pt]
Gautam (2020)  & India & $AOD$ & From this study, it was emerged an up-gradation of air quality in the Indian region just a week after lockdown restrictions took place, as a result of the sizeable reduction aerosol optical thickness concentration. \\
 \\  \addlinespace[3pt]
Agarwal et al (2020) & India and China & $NO_{2}$, $PM_{2.5}$ & It has been recorded a visible improvement in air quality parameters in some cities of India and China, selected on the basis of their availability of historical air pollution data, population density, monitoring station network, and the number of positive COVID$-$19 cases per million people.\\
 \\  \addlinespace[3pt]
Kanniah et al. (2020) & Malaysia and Southeast Asia & $AOD$, $PM_{10}$, $PM_{2.5}$, $NO_{2}$,$SO_{2}$, $CO$, $0_{3}$ & Over the Southeast Asia region, the analysis of air pollutants before and during the lockdown period  reported a significant drop in $AOD$, $PM_{2.5}$, $PM_{10}$, $NO_2$, $SO_2$, and $CO$. Besides, the authors reported $\sim$40 and $\sim$70 \% (in industrial and urban sites, respectively) reduction in $AOD$ level in Malaysia during March$-$April 2020 as compared to the same period in 2019 and 2018. \\
 \\  \addlinespace[3pt]
Kerimray et al. (2020) & Almaty (Kazakhstan) & $PM_{2.5}$, $CO$, $NO_{2}$, $0_{3}$, benzene, toluene & This study reported that in Almaty (Kazakhstan) $PM_{2.5}$ concentration reduced by 21\% with spatial variations of 6$-$34\% compared to the average of the same days in 2018$-$2019. $CO$ and $NO_2$ concentrations reduced by 49\% and 35\%, respectively while $O_3$ concentrations increased by 15\% compared to the preceding 17 days before the lockdown. Finally, concentrations of benzene and toluene were 2$-$3 times higher than in the same seasons of 2015$-$2019.\\
 \\  \addlinespace[3pt]
Zambrano-Monserrate  $\&$  Ruano (2020)  & Ecuador (Quito)  & $NO_{2}$, $PM_{2.5}$, $0_{3}$ & The quarantine policies adopted by the government of Ecuador have led to a   significant reduction of $NO_2$ and $PM_{2.5}$ concentrations. Specifically, it was found that the $NO_2$ concentrations of 2020 were, on average, 5.6 times less than the 2018 concentrations and 4.8 times less than those from 2019. Likewise, compared with these years, the $PM_{2.5}$ concentrations were 1.5 and 1.6 times lower, respectively. On the other hand, regarding $O_3$ concentrations, it was arisen that the ozone levels in 2020 were much higher than the levels in 2018 and 2019.\\
 \\  \addlinespace[3pt]
Dantas et al. (2020) & Rio de Janeiro (Brazil) & $NO_{2}$, $CO$, $O_{3}$, $NMHC$ & The authors showed how the reduction of  road traffic and economic activities led to the decrease in $CO$ and $NO_2$ levels and, by contrast, to the increase in ozone concentrations.\\
 \\  \addlinespace[3pt]
Nakata  $\&$ Urban (2020) & S\~{a}o Paulo state (Brazil) & $NO_{x}$, $NO_{2}$, $CO$, $O_{3}$  & Data from four air quality stations in S\~{a}o Paulo Brazil were analyzed to assess air pollutant concentrations variations during the partial lockdown. Overall, drastic reductions on $NO$ (up to$-$77.3\%), $NO_2$ (up to$-$54.3\%), and $CO$ (up to$-$64.8\%) concentrations were observed in the urban area during partial lockdown compared to the five-year monthly mean. By contrast, an increase of approximately 30\% in ozone concentrations was observed.\\

\end{longtable}
\end{center}

\end{document}